\documentclass[preprint,preprintnumbers,nofootinbib]{revtex4-1}

\usepackage[margin=20truemm]{geometry}
\usepackage{graphicx}   
\usepackage{dcolumn}    
\usepackage{subfigure}
\usepackage{bm}
\usepackage{amsmath,amssymb,amsfonts,amsthm}
\usepackage{latexsym}
\usepackage{bbm}        
\usepackage{color}
\usepackage{ulem}       
\usepackage{multirow}   
\usepackage{epsfig}
\usepackage{physics}    
\usepackage{comment}
\usepackage{appendix}
\usepackage{subcaption} 
\usepackage{xcolor}
\usepackage[justification=raggedright,singlelinecheck=false,labelfont=bf,labelsep=colon]{caption}

\usepackage{hyperref}
\hypersetup{
  setpagesize=false,
  bookmarksnumbered=true,
  bookmarksopen=true,
  colorlinks=true,
  linkcolor=blue,
  citecolor=red,
}

\begin{document}

\title{Effects of Early-Universe Inhomogeneity on Bubble Formation: Primordial Black Holes as an Extreme Case}

\author{Yijie Chang}
\email{changyj9@mail2.sysu.edu.cn}
\author{Haowen Deng}
\author{Yefeng Wang}
\author{Ran Ding}
\affiliation{School of Physics and Astronomy,
Sun Yat-sen University (Zhuhai Campus), Zhuhai 519082, China}

\author{Shihang Tang}
\email{tangshx5@mail2.sysu.edu.cn}
\author{Fa Peng Huang}
\email{Corresponding author. huangfp8@sysu.edu.cn}
\affiliation{MOE Key Laboratory of TianQin Mission,
TianQin Research Center for Gravitational Physics \& School of Physics and Astronomy,
Frontiers Science Center for TianQin,
Gravitational Wave Research Center of CNSA,
Sun Yat-sen University (Zhuhai Campus), Zhuhai 519082, China}

\bigskip

\date{\today}
\begin{abstract}
Our early Universe is not perfectly homogeneous and it may contain some inhomogeneous sources, which might distort the local spacetime and modify the bubble nucleation rate. 
Taking the primordial black hole as an extreme example, we investigate the bubble nucleation rate of a first-order phase transition in the vicinity of primordial black holes or other primordial gravitational sources. Our analysis reveals that the presence of primordial black holes can reduce the effective action and might modify the nucleation rate due to their gravitational effects, potentially altering the dynamics of the phase transition in the early universe and producing new gravitational wave signals since the gravitational effects of the primordial black hole or other possible inhomogeneous sources could lead to nucleation of non-spherical symmetric bubbles.
\end{abstract}

\maketitle

\section{Introduction}\label{sec:intro}

In particle cosmology, first-order phase transition (FOPT) dynamics play a central role in a number of important phenomena, including electroweak baryogenesis~\cite{Huang:2025cqi, White:2022ufa, vandeVis:2025efm, Morrissey:2012db}, the production of superheavy dark matter~\cite{witten:1984, Krylov:2013qe, Huang:2017kzu, Jiang:2023qbm, Baker:2020, Chway:2019kft, Azatov:2021ifm, Jiang:2024zrb}, the formation of primordial black holes (PBHs)~\cite{combinedPBHRefs}, the creation of primordial magnetic fields~\cite{Vachaspati:1991nm}, and the generation of stochastic gravitational waves~\cite{Huber:2008hg, Caprini:2015zlo, Hindmarsh:2017gnf, Qiu:2025tmn}. In the early Universe, a FOPT proceeds through the nucleation and expansion of bubbles, during which the metastable false vacuum decays into the true vacuum.

The bubble nucleation rate is the essential quantity for determining the phase transition dynamics and the related particle cosmology problems mentioned above. 
The pioneering studies of false vacuum decay were carried out by Coleman and Callan~\cite{PhysRevD.15.2929,PhysRevD.16.1762}, who analyzed the nucleation of $O(4)$-symmetric bubbles in flat spacetime at zero temperature~\footnote{In focusing exclusively on the gravitational enhancement induced by the inhomogeneity, we intentionally omit the consideration of recent advanced techniques~\cite{Jinno:2021, Hayashi:2021kro, Ai:2019fri} beyond the the Euclidean computation method~\cite{PhysRevD.15.2929,PhysRevD.16.1762} in this work. Direct approach of quantum tunneling in quantum field theory was studied in Refs.~\cite{Andreassen:2016cvx, PhysRevLett.117.231601}.}. Linde subsequently generalized this framework to finite temperature~\cite{LINDE198137,Linde:1981zj}, rendering it applicable to realistic early-Universe conditions. Coleman and collaborators also examined the gravitational backreaction of the nucleated bubble on the background spacetime~\cite{PhysRevD.21.3305}. The conventional analysis of the false vacuum decay is usually built upon the assumption of a highly symmetric and homogeneous background. 
The problem of vacuum decay in flat or homogeneous backgrounds has been extensively studied, establishing a standard paradigm for analyzing phase transitions in the early Universe.

However, the early universe is far from perfectly homogeneous or flat and it may contain some inhomogeneous sources such as PBHs~\cite{Hawking:1971ei, Carr:1974nx, Carr:1975qj} of various masses, monopoles, cosmic strings, and the domain wall. These inhomogeneous sources may significantly distort the local spacetime geometry and modify the bubble nucleation rate of a FOPT. In such cases, the $O(3)$ spatial symmetry of the bubble may be broken, with the bubble deformed. This deformation is expected to influence subsequent bubble collisions and, consequently, the resulting gravitational wave signatures. Since our primary observational probe of a FOPT comes from gravitational waves, understanding how bubble deformation arises and evolves is of crucial importance. Therefore, studying cosmological FOPTs in curved spacetime, with the existence of these inhomogeneous sources, is of great theoretical and phenomenological interest~\cite{BEREZIN1988397}, particularly for the physical processes involving FOPTs outlined in the opening paragraph.

Several works have discussed the bubble nucleation in the presence of topological defects—such as cosmic strings, monopoles, and domain walls—in various curved spacetime backgrounds~\cite{Oshita:2018ptr,Hiscock:1995ma,PhysRevD.47.2324,Dasgupta:1997kn,PhysRevD.35.3100,PhysRevD.34.1237,Kumar:2008jb,Firouzjahi:2020hfq,Steinhardt:1981ec,Lee:2013ega,Kumar:2010mv,Blasi:2022woz}. Extending beyond such configurations, a number of works have also considered black holes as potential nucleation sites~\cite{PhysRevD.32.1333,Cheung:2013sxa,Green:2006nv,Burda:2015yfa,Mukaida:2017bgd,El-Menoufi:2020ron}, typically within idealized settings such as Schwarzschild–de Sitter spacetime. These analyses suggest that black holes can substantially reduce the nucleation action, thereby modifying the vacuum decay rate~\cite{PhysRevD.35.1161,Ai:2018rnh,PhysRevD.106.125001,Gregory:2013hja}.
Motivated by these studies, we directly re-examine bubble nucleation in Schwarzschild geometry. Our calculations confirm that the presence of a Schwarzschild black hole significantly suppresses the nucleation action.

\begin{figure}[htbp]
    \centering
    \includegraphics[width=1\linewidth]{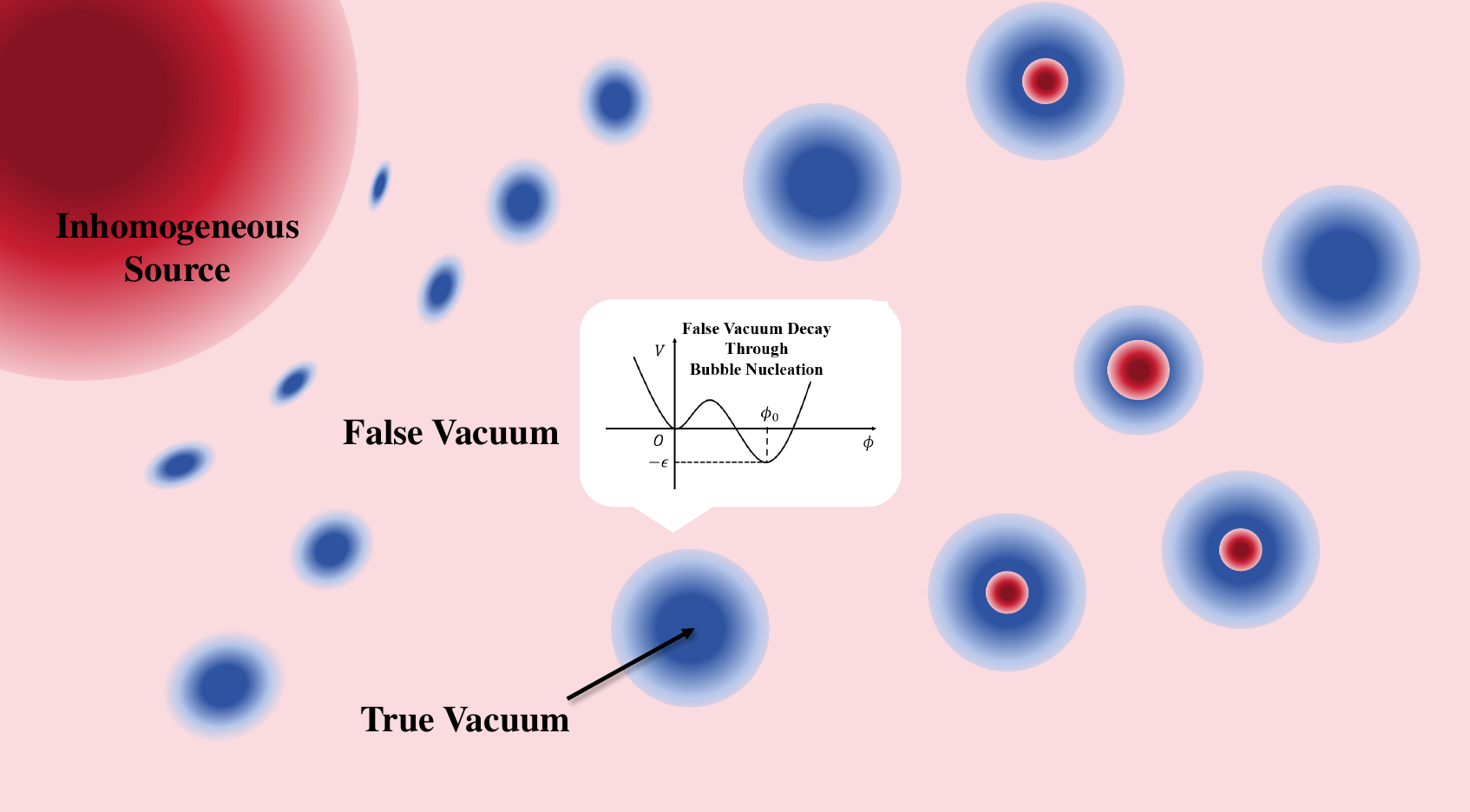}
    \caption{Bubbles nucleation at or around inhomogeneous sources. Here the deep red circles denote possible inhomogeneous sources, such as PBHs. The blue blobs depict the bubbles with or without deformation.}
    \label{fig:placeholder}
\end{figure}

Beyond the central-nucleation scenario, an important unresolved issue concerns bubbles that nucleate at finite distances from the black hole. Miyachi and Soda have examined such off-center configurations in a two-dimensional black hole spacetime, offering qualitative indications of how similar effects may arise in four dimensions~\cite{Miyachi:2021bwd}. However, a complete and self-consistent analysis of off-center nucleation in four-dimensional Schwarzschild spacetime is still lacking. In this work, we extend this line of research to the fully four-dimensional Schwarzschild geometry and systematically investigate off-center bubble nucleation. We study the bubble deformation due to the presence of a PBH, which may reduce the effective action of the bubble and enhance the nucleation rate under specific conditions. We also find that the deformed bubble profile could serve as a novel source of gravitational waves. A schematic illustration of the physical setup is presented in Fig.~\ref{fig:placeholder}.

This paper is organized as follows. In Sec.~\ref{sec:BBN_in_Flat}, we review the method developed by Coleman~\cite{PhysRevD.15.2929} and Linde~\cite{Linde:1981zj} for calculating the bubble nucleation rate in flat spacetime at finite temperature. In Sec.~\ref{sec:pbhcenter}, we first examine the nucleation of bubbles centered on a PBH. These bubbles possess $O(3)$ spatial symmetry, and our analysis confirms that the presence of a black hole enhances the nucleation rate relative to that in flat spacetime. In Sec.~\ref{sec:phbout}, we focus on the nucleation of bubbles outside a PBH. This section is divided into three parts. In Sec.~\ref{subsec:ballbubbleapprox}, we study bubble nucleation in regions far from the black hole, where the spacetime can be approximated as asymptotically flat. In this regime, the bubbles remain nearly spherical, and the calculation is relatively straightforward. In Sec.~\ref{subsec:ballbubblevalidity}, we examine the validity of the spherical bubble approximation introduced earlier. Our results indicate that the bubble largely maintains its spherical shape under small fluctuations. In Sec.~\ref{subsec:bubblenearpbh}, we then investigate bubble nucleation near the black hole, where gravitational effects are strong and the bubble shape becomes significantly deformed. Although the shape is nontrivial, the system preserves an $O(1)$ symmetry, which allows us to formulate the problem and numerically solve for the bubble profile from first principles. In Sec.~\ref{sec:discussion of gw}, we provide a discussion on the potential gravitational wave signals from deformed vacuum bubbles nucleated non-centrally around PBHs. Finally, a concise conclusion is given in Sec.~\ref{sec:sum}.

\section{bubble nucleation rate in flat spacetime at finite temperature}
\label{sec:BBN_in_Flat}

For the purposes of our subsequent analysis, a brief review of the bubble nucleation rate is presented in this section.
We commence with bubbles formed in Euclidean spacetime at zero temperature~\cite{PhysRevD.15.2929}.
In this case, the bubble possesses $O(4)$ symmetry. Therefore, its action can be written as
\begin{equation}
    S_4=\int d\tau d^3x\left[\frac{1}{2}\left(\frac{\partial\phi}{\partial\tau}\right)^2+\frac{1}{2}\left(\vec{\nabla}\phi\right)^2+V\right]=2\pi^2\int_0^\infty\rho^3d\rho\left[\frac{1}{2}\left(\frac{d\phi}{d\rho}\right)^2+V\right],
    \label{eq:S_4_ini}
\end{equation}
where the corresponding equation of motion for the scalar field $\phi$ satisfies:
\begin{equation}
    \frac{d^2\phi}{d\rho^2} + \frac{3}{\rho} \frac{d\phi}{d\rho} = V'(\phi).
    \label{eqS4}
\end{equation}

Here $\rho\equiv \left(\tau^2+|\vec{x}|^2\right)^{1/2}$.
 The potential is defined with $V=0$ for the false vacuum and $V=-\epsilon<0$ for the true vacuum, with a smooth interpolating path for $V$ between the two vacua.

For the thin-wall approximation, Eq.~\eqref{eqS4} can be simplified :
\begin{equation}
    \frac{d^2\phi}{d\rho^2} = V'(\phi).
\end{equation}

Besides, we assume that the scalar field $\phi$ remains constant in false vacuum and true vacuum except the phase transition regime.
Thus, the total action for a bubble in the thin-wall approximation is
\begin{equation}
    S_4=-\epsilon\int_0^R2\pi^2\rho^3d\rho+2\pi^2 R^3\int_{\rho\approx R}d\rho \left[ \frac{1}{2} \left( \frac{d\phi}{d\rho} \right)^2 + V(\phi) \right]=-\epsilon V^{(4)}_{\text{bubble}}+S_1S^{(3)}_{\text{bubble}},
    \label{S_4inflat}
\end{equation}
where $R$ is the radius for the bubble, $V^{(4)}_{\text{bubble}}$ denotes the four-dimensional bubble volume, and $S^{(3)}_{\text{bubble}}$ represents the four-dimensional surface area. 
 $S_1$ can be regarded as the one-dimensional action:
\begin{equation}
    S_1=\int_{\rho \approx R} d\rho \left[ \frac{1}{2} \left( \frac{d\phi}{d\rho} \right)^2 + V(\phi) \right].
\end{equation}

Then we can calculate the bubble nucleation rate at zero temperature as follows:
\begin{equation}
    \Gamma\propto e^{-S_4}.
\end{equation}

With the zero-temperature formalism developed above, one can naturally generalize the bubble nucleation at high temperature~\cite{Linde:1981zj}, which is essential to study the phase transition physics in the early universe. At high temperature, one should use the finite-temperature quantum field theory which predicts
the periodicity of the bubble configuration along the Euclidean time direction, namely, $\phi(\tau,\vec{x})=\phi(\tau+1/T,\vec{x})$. The period is just the inverse of the temperature with the conventional definition $\beta=T^{-1}$. Consequently, at very high temperature, the period $\beta$ is small, hence one can have the approximation that $\left(\frac{\partial\phi}{\partial\tau}\right)^2=0$. 
Now the bubble only remains its
$O(3)$ symmetry in space~\cite{Linde:1981zj}. 
Therefore, Eq.~\eqref{eq:S_4_ini} can be simplified and factorized as:
\begin{equation}
\begin{split}
    		S_4 =& \int_0^\beta d\tau  \int d^3 x \left[ \frac{1}{2} \left( \vec{\nabla} \phi \right)^2 + V(\phi, T) \right] \\
            =&\beta\cdot\int d^3 x \left[ \frac{1}{2} \left( \vec{\nabla} \phi \right)^2 + V(\phi, T) \right] \\
            \equiv& \frac{S_3}{T},
\end{split}
\end{equation}
where the action of the spatial part $	S_3 = \int d^3 x \left[ \frac{1}{2} \left( \vec{\nabla} \phi \right)^2 + V(\phi, T) \right]$. Here it's worth noticing that at high temperature the potential $V(\phi)$ should be replaced with the finite-temperature effective potential $V(\phi,T)$.

Therefore, following a similar analysis as above and by exploiting the $O(3)$ symmetry of the bubble, we can also write the $S_3$ as
\begin{equation}
    S_3 = \int d^3 x \left[ \frac{1}{2} \left( \vec{\nabla} \phi \right)^2 + V(\phi, T) \right]=4\pi\int_{0}^{\infty}\,r^2dr\left[\frac{1}{2}\left(\frac{d\phi}{dr}\right)^2+V(\phi, T)\right],
\end{equation}
and the corresponding equation of motion of $\phi$ in $O(3)$ case is 
\begin{equation}
    \frac{d^2\phi}{dr^2} + \frac{2}{r} \frac{d\phi}{dr} = \frac{dV(\phi, T)}{d\phi}.
\end{equation}

Similarly, under the thin-wall approximation, we can also readily obtain the total action of this $O(3)$ bubble
\begin{equation}
    S_4=\frac{S_3}{T}=\frac{-\epsilon V^{(3)}_{\text{bubble}}+S_1S^{(2)}_{\text{bubble}}}{T},
\end{equation}
where $V^{(3)}_{\text{bubble}}$ denotes the three-dimensional bubble volume, and $S^{(2)}_{\text{bubble}}$ represents the three-dimensional surface area. 

The final bubble nucleation rate at high temperature is obtained as
\begin{equation}
    \Gamma\propto e^{-S_4}=e^{-\frac{S_3}{T}}.
\end{equation}

After including the prefactor~\cite{Linde:1981zj}, $\Gamma$ can be approximately expressed as:
\begin{equation}
    \Gamma\sim T^4\left(\frac{S_3}{2\pi T}\right)^{\frac{3}{2}}e^{-\frac{S_3}{T}}.
    \label{concise_form_of_Gamma}
\end{equation}

\section{bubble nucleation rate with the black hole at its center}\label{sec:pbhcenter}

In order to investigate the effects of inhomogeneous sources, we extend the framework developed above to curved spacetime. Taking the Schwarzschild black hole as an example, after performing a Wick rotation~\cite{Gibbons:1976ue}, the Euclidean Schwarzschild metric can be written as
\begin{equation}
    ds^2 = (1 - 2GM/r)d\tau^2 + (1 - 2GM/r)^{-1}dr^2 + r^2 d\Omega^2.
    \label{eq:ds**2}
\end{equation}

The Euclidean action of the scalar field can be expressed as
\begin{equation}
    S_4 = \int d^4x \sqrt{g} \left( \frac{R_S}{16\pi} + \mathcal{L} \right) 
    = \int d^4x \sqrt{g} \left( \frac{1}{2} g^{\mu\nu} \partial_\mu \phi \partial_\nu \phi + V(\phi) \right),
\end{equation}
where $R_S$ denotes the Ricci scalar of the Schwarzschild metric, to distinguish it from the bubble radius $R$.

To ensure a clear discussion and general conclusions, we consider a toy model with the following finite-temperature effective potential~\cite{Linde:1981zj}:
\begin{equation}
	V(\phi,T) = \frac{1}{2}\gamma (T^2 - T_c^2)\phi^2 - \frac{1}{3}\alpha T \phi^3 + \frac{1}{4}\lambda \phi^4,
	\label{eq:V_eff}
\end{equation}
where $\gamma$, $T_c$, $\alpha$ and $\lambda$ are constant parameters. We list benchmark values of these parameters which can produce FOPTs through bubble nucleation in Tab.~\ref{table:refrence}~\cite{Wang:2020jsg}.

\begin{table}[t]
	\centering
	\begin{tabular}{lccc}
    	\hline\hline
        \multicolumn{4}{c}{\textbf{Benchmark parameters for the finite-temperature effective potential}}\\
        \hline
		Parameter & Value & Description \\ 
		\hline
			$\gamma$ & $1.5$ & Quadratic term coefficient in $V(\phi, T)$ \\
			$\alpha$ & $0.402$ & Cubic term coefficient in $V(\phi, T)$ \\
			$\lambda$ & $0.1$ & Quartic term coefficient in $V(\phi, T)$ \\
			$T_* \simeq T_c$ & $63.5169\,\mathrm{GeV}$ & Phase transition temperature  \\
			$G$ & $6.7\times10^{-39}\,\mathrm{GeV}^{-2}$ & Newton’s constant  \\
			$M$ & $1.0\times10^{40}\,\mathrm{GeV}$ & Black hole mass  \\
		\hline\hline
	\end{tabular}
    \caption{Benchmark Parameters used in the numerical simulations.}
    \label{table:refrence}
\end{table}

We define $\phi_0$ and $T_0$ as the field value and temperature at which $V(\phi, T)$ has two degenerate minima, i.e.,
\begin{equation}
    V(0, T_0) = V(\phi_0, T_0).
\end{equation}

At this point, the effective potential can be approximated by
\begin{equation}
	V(\phi) = \frac{1}{4} \lambda \phi^2 (\phi_0 - \phi)^2,
	\label{eq:V_appro}
\end{equation}
and the energy density difference between the false vacuum and true vacuum satisfies $\epsilon = 0$. Substituting Eq.~\eqref{eq:V_appro} into  Eq.~\eqref{eq:V_eff} one obtains 
\begin{equation}
	\phi_0 = \frac{2\alpha T_0}{3\lambda},
	\label{eq:phi_0}
\end{equation}
and
\begin{equation}
	\left(1 - \frac{2\alpha^2}{9\lambda\gamma}\right) T_0^2 = T_c^2.
	\label{eq:T_0}
\end{equation}

To ensure consistency with the approximation that the phase transition temperature $T_* \simeq T_c$, we require $(T_0 - T_c)/T_c \ll 1$. Under this assumption, the energy density difference $\epsilon$ can be approximated by expanding the potential around $(\phi_0, T_0)$:
\begin{equation}
      -\epsilon(T) = V(\phi_0, T) - 0 \approx (T - T_0)\frac{\partial V}{\partial T}\Big|_{\phi=\phi_0,\,T=T_0}   = -\frac{4 T_0 T_c^2 \alpha^2 \gamma}{9 \lambda^2} (T_0 - T).
    \label{eq:epsilon}
\end{equation}

Since the Schwarzschild spacetime is spherically symmetric, a bubble nucleated at the center of the black hole naturally exhibits $O(3)$ symmetry, with $\phi = \phi(r)$. Because the interior of the black hole has a highly nontrivial causal structure, we consider only the case where the bubble radius is much larger than the event horizon radius, allowing us to neglect the contribution from the region inside the horizon.

We assume that the true vacuum occupies the interior of the bubble, while the exterior remains in the false vacuum, with quantum tunneling occurring across the boundary. The scalar field configuration is then
\begin{align}
    &r < R:&\phi = \phi_0, \quad &V(\phi_0) = -\epsilon, \\
    &r \approx R:&\phi = \phi_1(r - R), \quad &V(\phi_1) = \frac{\lambda}{4} \phi_1^2 (\phi_0 - \phi_1)^2, \\
    &r > R:&\phi = 0, \quad &V(\phi=0) = 0.
\end{align}

The Klein–Gordon equation governing the tunneling field $\phi_1$ reads
\begin{equation}
	(1 - 2GM/r)\frac{d^2\phi_1}{dr^2} + \frac{2r - 2GM}{r^2}\frac{d\phi_1}{dr} = \frac{dV}{d\phi_1}.
\end{equation}

We simplify this equation by expanding $(1 - 2GM/r)$ around $r = R$ and applying the thin-wall approximation, leading to:
\begin{equation}
	(1 - 2GM/R)\frac{d^2\phi_1}{dr^2} = \frac{dV}{d\phi_1}.
\end{equation}

Defining  $\tilde{V} =V/(1-2GM/R)$, the equation takes the well-known form of a soliton equation~\cite{PhysRevD.11.1486}. Recognizing this form immediately yields the solution
\begin{equation}
    \frac{d\phi_1}{dr} = \sqrt{\frac{2V}{1 - 2GM/R}}.
\end{equation}

The total Euclidean action is then
\begin{equation}
\begin{split}
	S_4 &= \int d^4x \sqrt{g} \left(\frac{1}{2} g^{\mu\nu}\partial_\mu\phi\partial_\nu\phi + V(\phi)\right) \\
    &= \frac{1}{T} \int r^2 \sin\theta \, dr \, d\theta \, d\varphi \left(\frac{1}{2} g^{rr}\partial_r\phi \partial_r\phi + V(\phi)\right) \\
    &= \frac{1}{T}\left[-\epsilon\frac{4\pi R^3}{3} + 4\pi R^2 \int_{r\approx R} dr \left(\frac{1}{2} g^{rr}\partial_r\phi_1 \partial_r\phi_1 + V(\phi_1)\right)\right].
\end{split}
\end{equation}

Defining
\begin{equation}
\begin{split}
    S_1 =& \int_{r\approx R} dr \left(\frac{1}{2} g^{rr}\partial_r\phi_1 \partial_r\phi_1 + V(\phi_1)\right)\\
    =&\int_0^{\phi_0} d\phi_1\, \sqrt{2V(\phi_1)\left(1 - \frac{2GM}{R}\right)}\\
    \equiv& k\sqrt{1 - 2GM/R},
    \label{eq:S1}
\end{split}    
\end{equation}
where $k \equiv \frac{2^{3/2}\alpha^3}{3^4\lambda^{5/2}} T_c^3$, the total Euclidean action becomes
\begin{equation}
    S_4 = \frac{1}{T} \left(-\epsilon \frac{4\pi R^3}{3} + 4\pi k R^2 \sqrt{1 - 2GM/R}\right).
    \label{S_4atBHcenter}
\end{equation}

By analogy with the definition of the volume energy density $\epsilon$, we recognize $S_1$ as the surface energy density of the tunneling interface.

Minimizing the action, $\delta S_4/\delta R = 0$, yields the critical bubble radius:
\begin{equation}
	-4\pi R^2\epsilon + 8\pi k R\sqrt{1 - 2GM/R} + \frac{4\pi kGM}{\sqrt{1 - 2GM/R}} = 0.
\end{equation}

\begin{figure}[h!]
    \centering
    \includegraphics[width=1\linewidth]{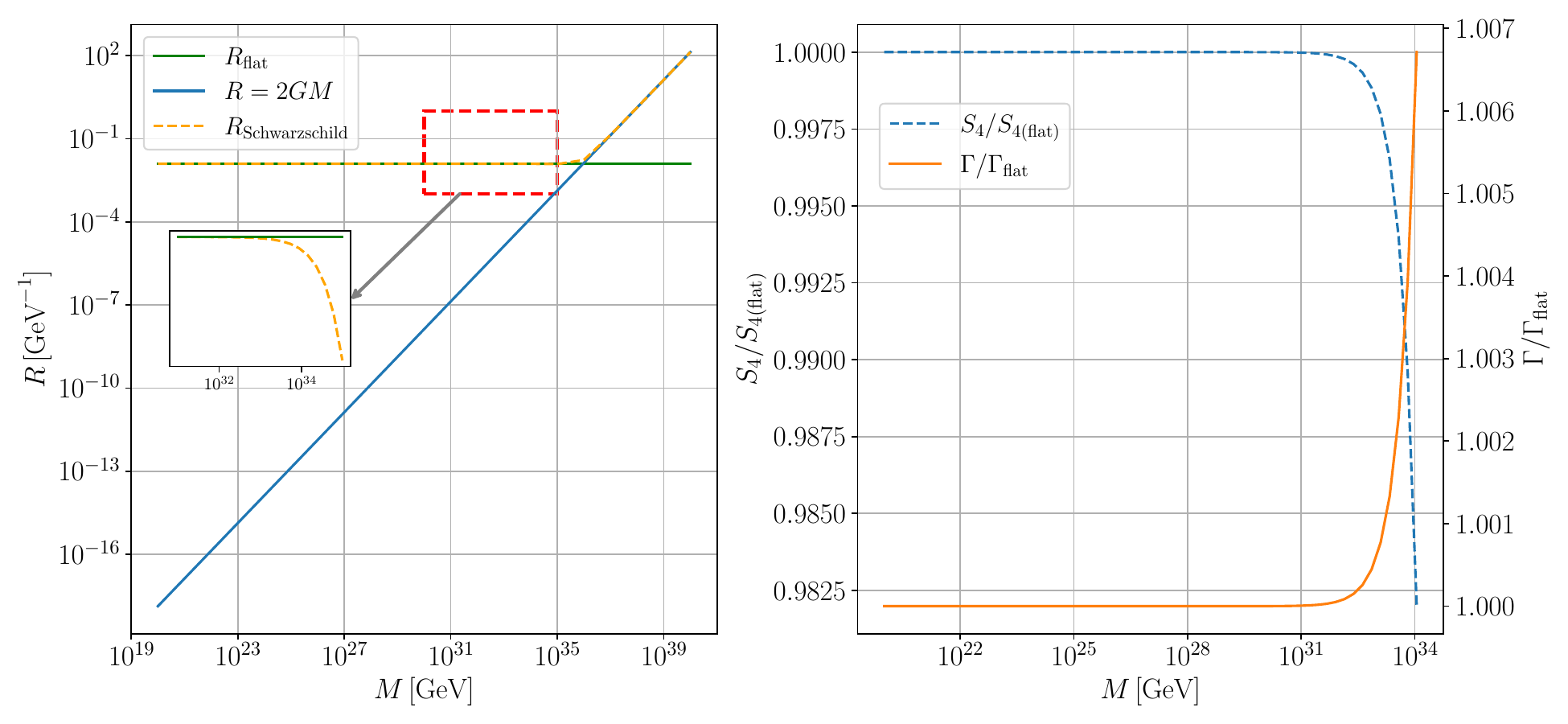}
    \caption{Left panel: Critical bubble radius as functions of black hole mass; Right panel: Corresponding action and nucleation rate as functions of black hole mass normalized to their Euclidean values for comparison. In the left panel, the green line represents the critical bubble radius \( R_{\text{flat}} \) in flat spacetime, the orange dashed curve depicts the critical radius \( R_{\text{Schwarzschild}} \) in Schwarzschild spacetime, and the blue line indicates the horizon at \( R = 2~GM \); in the right panel, the blue dashed shows the normalized action ratio while the yellow curve gives the nucleation rate ratio.}
    \label{R-GM}
\end{figure}

This equation has only one physical solution, which always satisfies $R > 2GM$, implying that a bubble centered on the black hole does not nucleate within the horizon. Using the parameters listed in Table~\ref{table:refrence}, we numerically determine the critical bubble radius as a function of the black hole mass in the range $10^{20}\text{--}10^{40}\,\mathrm{GeV}$. For $M \gtrsim 10^{34}\,\mathrm{GeV}$, the resulting bubble radius becomes comparable to the black hole horizon, and the influence of the spacetime inside the horizon can no longer be neglected. We therefore restrict our analysis to $M \lesssim 10^{34}\,\mathrm{GeV}$, where the approximation that the bubble radius should be much larger than the horizon of the black hole remains valid. 

Within this regime, we use Eqs.~\eqref{S_4atBHcenter} and~\eqref{concise_form_of_Gamma} to evaluate the bubble action and nucleation rate, and present the results—including the bubble radius—as functions of black hole mass in Fig.~\ref{R-GM}. The action and nucleation rate are normalized to the corresponding values in Euclidean flat spacetime for comparison. The left panel illustrates the radius behavior: the green line represents the critical bubble radius \( R_{\text{flat}} \) in flat spacetime, the orange dashed curve depicts the critical radius \( R_{\text{Schwarzschild}} \) in Schwarzschild spacetime, and the blue line indicates the horizon at \( R = 2~GM \), clearly showing where the bubble radius substantially exceeds the black hole size. The right panel displays the normalized action ratio \( S_4/S_{4\text{(flat)}} \) (blue dashed) and nucleation rate ratio \( \Gamma/\Gamma_{\text{flat}} \) (yellow). As shown, the reduction in bubble radius accordingly leads to a lower nucleation action, which in turn results in a higher nucleation rate, demonstrating that black holes effectively catalyze vacuum bubble nucleation.

\section{Bubble nucleation rate outside of black hole}\label{sec:phbout}

In this section, we focus on the insufficiently studied scenario where bubble nucleation occurs outside the black hole horizon. This configuration, where the critical bubble's center lies at a finite distance from the black hole, requires dedicated theoretical treatment.

\subsection{Bubble nucleation rate far from the black hole}
\label{subsec:ballbubbleapprox}

To analyze this scenario, we first consider a small local region at a distance $r_0$ outside a Schwarzschild black hole, where the spacetime can be approximately treated as locally flat. In this region the critical bubble radius $R$ is assumed to be much smaller than the distance $r_0$ and the local line element thus can be written as
\begin{equation}
    ds^2 = \left(1 - \frac{2GM}{r_0}\right)d\tau^2 + \left(1 - \frac{2GM}{r_0}\right)^{-1}dr^2 + r^2 d\Omega^2.
\end{equation}

Assuming that the scalar field $\phi$ depends only on $r$, therefore the Euclidean action can then be correspondingly simplified to
\begin{equation}
    S_4 = \frac{4\pi}{T}\int r^2 \left[\frac{1}{2}\left(1 - \frac{2GM}{r_0}\right)\left(\frac{d\phi}{dr}\right)^2 + V(\phi)\right]dr,
\end{equation}
and the corresponding equation of motion with the thin-wall approximation takes the form
\begin{equation}
    \left(1 - \frac{2GM}{r_0}\right)\frac{d^2\phi}{dr^2} = V'(\phi).
\end{equation}

Its solution is the well-known soliton solution~\cite{PhysRevD.11.1486}; hence, one can readily derive
\begin{equation}
    \frac{d\phi}{dr} = \sqrt{\frac{2V(\phi)}{1 - 2GM/r_0}}.
\end{equation}

Owing to the fact that inside the bubble ($r<R$) the field takes $\phi = \phi_0$ with $V(\phi) = -\epsilon$, while near the bubble wall ($r \approx R$) the field exhibits rapid variation, the action can be correspondingly simplified to
\begin{equation}
    S_4 = \frac{4\pi}{T}\left[\frac{1}{3}R^3(-\epsilon) + R^2 S_1\right].
\end{equation}

Upon minimizing the action $S_4$ with respect to the bubble radius $R$, one find the critical radius
\begin{equation}
    R = \frac{2S_1}{\epsilon},
\end{equation}
and the corresponding action takes the form
\begin{equation}
    S_4 = \frac{16\pi S_1^3}{3T\epsilon^2}.
\end{equation}

The surface energy density $S_1$ can be further expressed as
\begin{equation}
    \begin{split}
        S_1 &= \int_{r\approx R} dr \left[\frac{1}{2}\left(1 - \frac{2GM}{r_0}\right)\left(\frac{d\phi}{dr}\right)^2 + V(\phi)\right] \\
            &= \int_0^{\phi_0} d\phi\, \sqrt{2V(\phi)\left(1 - \frac{2GM}{r_0}\right)},
    \end{split}
\end{equation}
and by substituting the effective potential given in Eqs.~(\ref{eq:V_eff})-(\ref{eq:epsilon}),
we obtain
\begin{equation}
    S_1 = \sqrt{2\lambda\left(1 - \frac{2GM}{r_0}\right)}\,\frac{\phi_0^3}{12}
        = \frac{2^{3/2}\alpha^3}{3^4\lambda^{5/2}}T_c^3 \sqrt{1 - \frac{2GM}{r_0}}.
    \label{eq:S1_toy}
\end{equation}

Using this expression, the critical bubble radius as a function of temperature and position is given by
\begin{equation}
    R(T, r_0) = \frac{2S_1}{\epsilon}
    = \frac{\sqrt{2}\alpha}{9\sqrt{\lambda}\gamma}\sqrt{1 - \frac{2GM}{r_0}}
      \frac{T_c}{T_0(T_0 - T)}.
    \label{eq:R_ball}
\end{equation}

Substituting Eq.~\eqref{eq:R_ball} into the action yields the corresponding Euclidean action
\begin{equation}
    S_4(T, r_0) =
    \frac{2^{9/2}\pi \alpha^5}{3^9 \lambda^{7/2}\gamma^2}
    \left(1 - \frac{2GM}{r_0}\right)^{3/2}
    \frac{T_c^5}{T_0^2 (T_0 - T)^2 T}.
    \label{eq:S4_ball}
\end{equation}

Consequently, according to Eq.~\eqref{concise_form_of_Gamma}, the bubble nucleation rate in the Schwarzschild background, under the leading-order approximation, is taking the form
\begin{equation}
    \Gamma(T, r_0) \sim T^4\left(\frac{S_4(T,\,r_0)}{2\pi}\right)^{3/2}\exp\left[-S_4(T,\,r_0)\right].
    \label{eq:Gamma_ball}
\end{equation}

The results from Eqs.~\eqref{eq:R_ball}-\eqref{eq:Gamma_ball} are summarized in Fig.~\ref{D-rGamma-r}. The left panel displays critical bubble diameters: the orange line (\(D_{\text{flat}}\)) serves as a constant flat-spacetime reference, while the bright cyan line (\(D_{\text{spherical}}\)) corresponds to spherical bubbles far from the black hole, lying below the reference at finite \(r_t\) and approaching it as \(r_t\) increases. The right panel shows corresponding action and nucleation rate ratios relative to the values in Euclidean flat spacetime. The purple curve (\(S_{4\text{(spherical)}}/S_{4\text{(flat)}}\)) remains below unity, decreasing sharply at small \(r_t\) and recovering to unity as \(r_t\) grows. The bright cyan curve (\(\Gamma_{\text{spherical}}/\Gamma_{\text{flat}}\)) increases from below to above unity as \(r_t\) rises to moderate values, eventually converging to unity at large \(r_t\).

These behaviors indicate that black-hole gravity substantially reduces the Euclidean action, thereby lifting the exponential suppression. Moderate gravitational fields enhance the nucleation rate, while both action and rate recover their flat-space values in the weak-field limit. Notably, for very small \(r_t\) (strong gravity), the nucleation rate remains suppressed relative to flat space despite the action reduction, likely due to the limited validity of Eq.~\eqref{eq:Gamma_ball} in such extreme regimes.

Overall, in the presence of a black hole, both the critical bubble radius and the Euclidean action are reduced, promoting nucleation in moderate fields. This reduction in both radius and action strengthens with larger black hole mass or nucleation nearer to the horizon. For small black holes or bubbles far from the horizon, the results smoothly approach those in flat spacetime~\cite{LINDE198137}.

\subsection{Validity of the spherical bubble approximation}
\label{subsec:ballbubblevalidity}

We first examine whether the assumption is valid:
\begin{equation}
    \frac{R(T_c)}{r_0} = 
    \frac{\sqrt{2}\alpha}{9\sqrt{\lambda}\gamma}
    \sqrt{\frac{1}{r_0^2} - \frac{2GM}{r_0^3}}
    \frac{T_c^2}{T_0^2(T_0 - T_c)} \ll 1.
\end{equation}

\begin{figure}[h]
    \centering
    \includegraphics[width=1.0\textwidth]{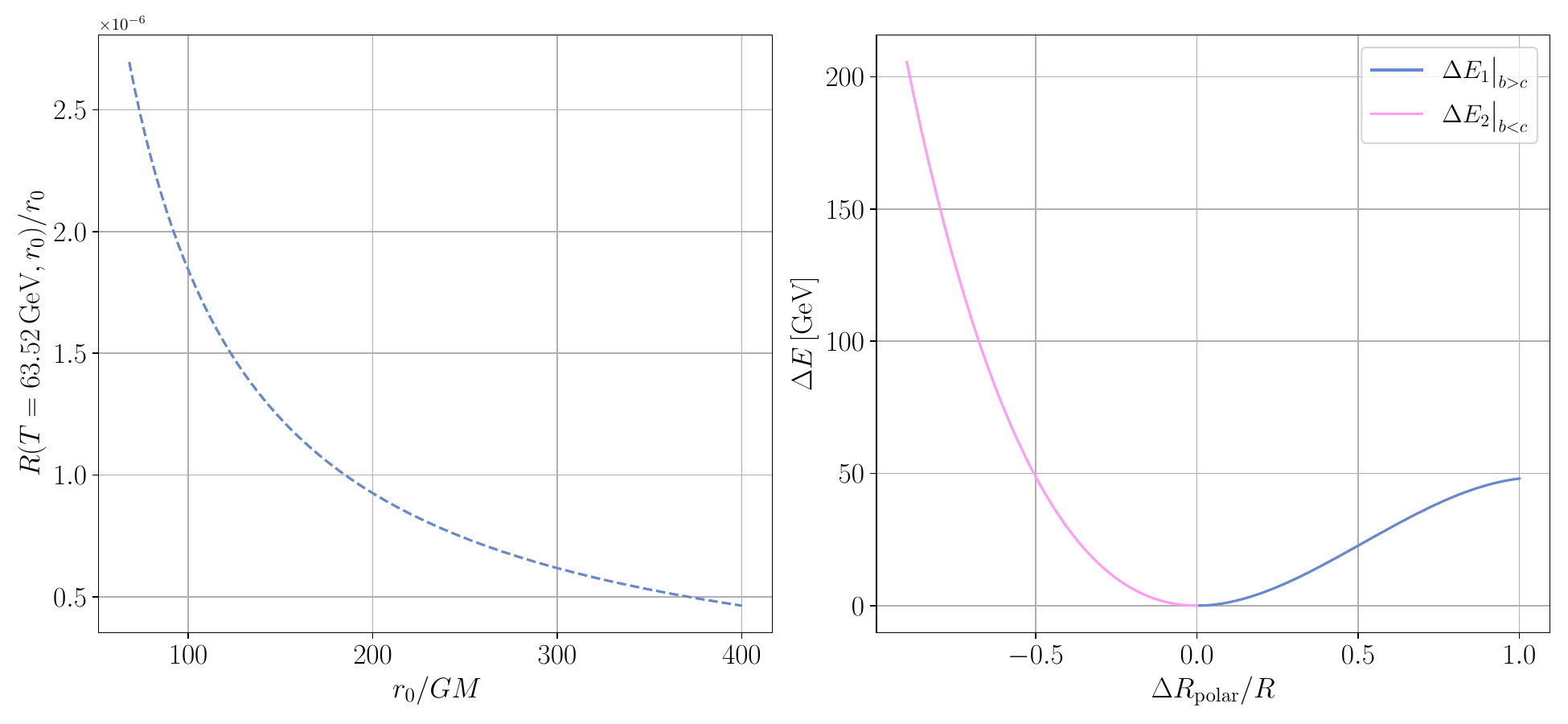}
    \caption{
    Left panel: Variation of critical bubble radius-to-distance ratio with respect to the distance from the black hole.
    Right panel: Total energy variation of a bubble nucleated at  $r=80~GM$ under a tidal deformation assumption
    $-2\Delta R_{\text{equatorial}} = \Delta R_{\text{polar}}$, where the left pink curve corresponds to the prolate spheroidal bubble, and the right blue curve to the oblate spheroidal bubble.
    }
    \label{fig:ball_Gamma_R}
\end{figure}

As shown in the left panel of Fig.~\ref{fig:ball_Gamma_R}, the ratio $R/r_0$ (blue dashed curve) remains on the order of $10^{-6}$, confirming the validity of the assumption $R \ll r_0$.

Moreover, we further re-discuss the stability of the spherical solution under small gravitational perturbation. To address this, we introduce a small tidal deformation to the bubble, which remains much smaller than the bubble’s characteristic size. To quantify this deformation, we expand the Newtonian potential $\Phi = -GM/|\vec{r}|$ at $r_0$ up to the second order in $\vec{R}$, where $\vec{r} = \vec{r_0} + \vec{R}$, yielding the tidal potential per unit mass:
\begin{equation}
    \Phi_{\text{tidal}} = \frac{GM}{2r_0^3}\left(R_x^2 + R_y^2 - 2R_z^2\right),
\end{equation}
where $\vec{r_0} = (0, 0, r_0)$ and $\vec{R} = (R_x, R_y, R_z)$.
To derive the effective mass density of the bubble, we employ the mass–energy relation, considering both the volume contribution $E_V = \frac{4\pi}{3}R^3(-\epsilon) = mc^2$ and the surface contribution $E_S = 4\pi R^2 S_1 = mc^2$, giving
\begin{equation}
    \rho_V = -\epsilon, \qquad \rho_S = S_1.
\end{equation}
Assuming that the initially spherical bubble is deformed tidally into an ellipsoid with polar axis $c$ and equatorial axis $b$, we parameterize
\begin{equation}
    R_x = b\gamma\sin\theta\cos\varphi, \quad
    R_y = b\gamma\sin\theta\sin\varphi, \quad
    R_z = c\gamma\cos\theta.
\end{equation}

Therefore, the tidal potential energy of this ellipsoidal bubble is then
\begin{equation}
    \begin{split}
        E_{\text{tidal}} &= \int \rho\,\Phi_{\text{tidal}}\,dV^{(3)}_{\text{bubble}} \\
        &= \int(-\epsilon)\frac{GM}{2r_0^3}(R_x^2 + R_y^2 - 2R_z^2)dV^{(3)}_{\text{bubble}}
           + \int S_1\frac{GM}{2r_0^3}(R_x^2 + R_y^2 - 2R_z^2)dS^{(2)}_{\text{bubble}},
    \end{split}
\end{equation}
where $dV^{(3)}_{\text{bubble}} = b^2c\gamma\sin\theta\,d\gamma\,d\theta\,d\varphi$ and
$dS^{(2)}_{\text{bubble}} = b\sin\theta\sqrt{c^2\sin^2\theta + b^2\cos^2\theta}\,d\theta\,d\varphi$.

After performing the integrations, we obtain
\begin{equation}
    E_{\text{tidal}} = \frac{4\pi(-\epsilon)GM}{15r_0^3}b^2c(b^2 - c^2) + \frac{S_1GM}{2r_0^3} I.
\end{equation}
Here, $I$ denotes the elliptic surface integral for the bubble, which, via the substitution $u\equiv\cos{\theta}$, takes the form:
\begin{equation}
    \begin{split}
        I &= 4\pi b \int_0^1 [b^2 - (b^2 + 2c^2)u^2] \sqrt{c^2 - (c^2 - b^2)u^2}\,du .
    \end{split}
\end{equation}

For different geometrical configurations of the elliptic bubble, $I$ has distinct analytic forms. In particular, for a prolate spheroidal bubble $(c>b)$ and an oblate spheroidal bubble $(c<b)$, the corresponding expressions must be evaluated separately.

For a prolate spheroid, where the polar axis $c$ exceeds the equatorial axis $b~(c>b)$, the integral $I_1$ becomes:
\begin{equation}
\begin{split}
    I_1=&-\frac{b\pi}{2(c^2-b^2)^{3/2}}\Bigg[(2b^5-7b^3c^2+2bc^4)\sqrt{c^2-b^2}\\
    &+(4b^4c^2-3b^2c^4+2c^6)\text{arccsc}(\frac{c}{\sqrt{c^2-b^2}})\Bigg].
\end{split}
\end{equation}

Using this result, the total variation in the bubble energy, $\Delta E_{\text{total1}}$, can be expressed as
\begin{equation}
    \begin{split}
            \Delta E_{\text{total1}}=&\left[\frac{4\pi(-\epsilon)GM}{15r_0^3}b^2c(b^2-c^2)+\frac{S_1GM}{2r_0^3}\cdot I_1\right]\\
            &+\left(\frac{4}{3}\pi cb^2-\frac{4}{3}\pi R^2\right)\cdot(-\epsilon)\\
            &+\left[2\pi b^2+\frac{2\pi c^2b}{\sqrt{c^2-b^2}}\arcsin(\frac{\sqrt{c^2-b^2}}{c})-4\pi R^2\right]\cdot S_1\,.
            \label{eq:Delta_E1}
    \end{split}
\end{equation}

Conversely, when the polar axis is shorter than the equatorial axis $(c<b)$, corresponding to a oblate spheroid, the integral $I_2$ is given by
\begin{equation}
\begin{split}
     I_2 =& \frac{b\pi}{2(b^2 - c^2)^{3/2}}\Bigg[
    (2b^4 - 7b^2c^2 + 2c^4)b\sqrt{b^2 - c^2}\\
    &+ c^2(4b^4 - 3b^2c^2 + 2c^4)\text{arccsch}\!\left(\frac{c}{\sqrt{b^2 - c^2}}\right)
      \Bigg]\,.
\end{split}
\end{equation}

In this case, the total bubble energy variation, $\Delta E_{\text{total2}}$, takes the form shown in
\begin{equation}
    \begin{split}
            \Delta E_{\text{total2}}=&\left[\frac{4\pi(-\epsilon)GM}{15r_0^3}b^2c(b^2-c^2)+\frac{S_1GM}{2r_0^3}\cdot I_2\right]\\
            &+\left(\frac{4}{3}\pi cb^2-\frac{4}{3}\pi R^2\right)\cdot(-\epsilon)\\
            &+\left[2\pi b^2+\frac{\pi b^2(1-e^2)}{e}\cdot \ln(\frac{1+e}{1-e})-4\pi R^2\right]\cdot S_1\,,
            \label{eq:Delta_E2}
    \end{split}
\end{equation}
where $e = \sqrt{b^2 - c^2}/b$ is the ellipticity.

\begin{figure}[h]
    \centering
    \includegraphics[width=0.8\textwidth]{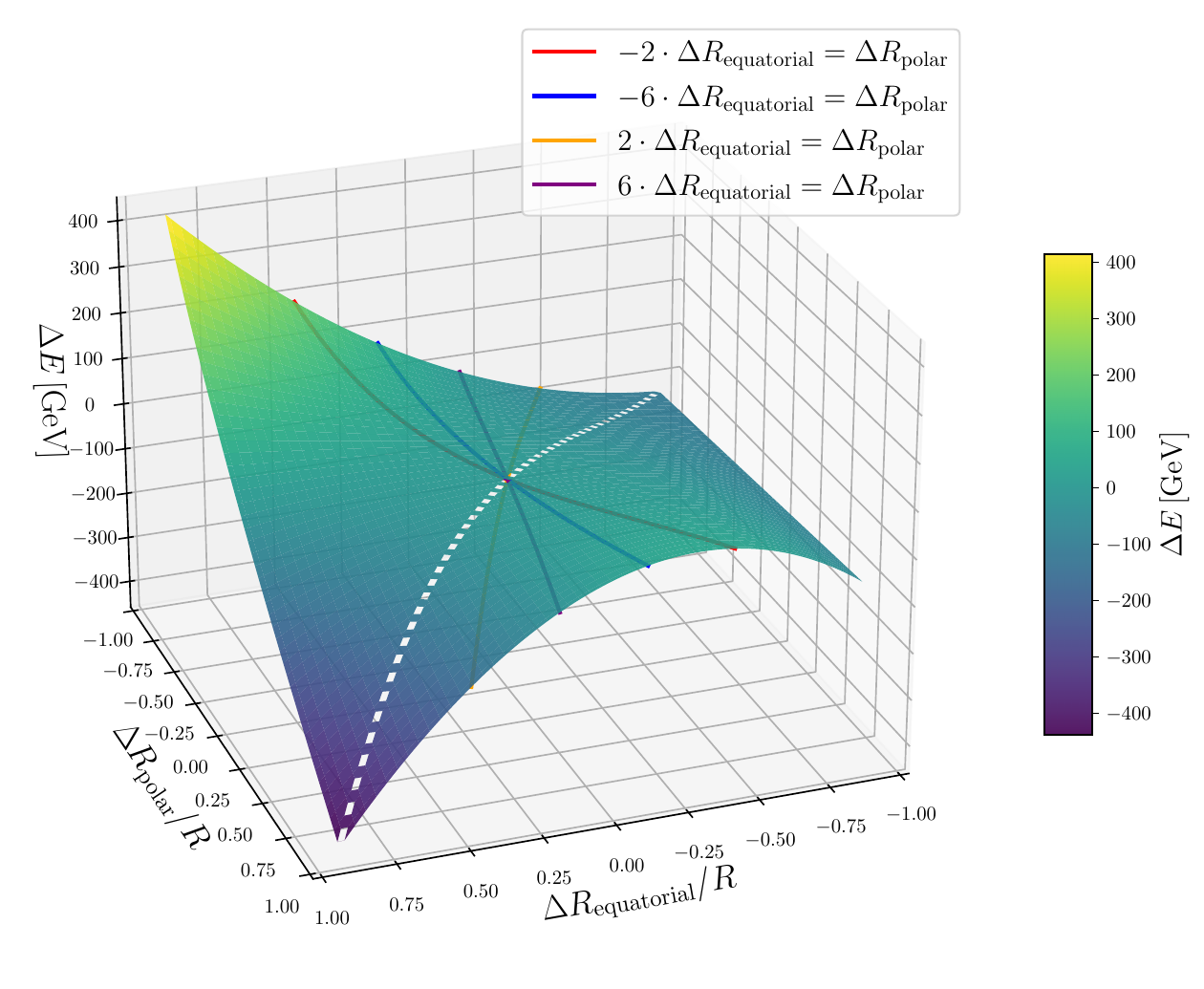}
    \caption{
    Landscape of the bubble energy difference as a function of longitudinal $R_{\text{equatorial}}$ (\(b\)) and transverse $R_{\text{polar}}$ (\(c\)) deformations, computed for a nucleation center at \(r = 80~GM\), with specific deformation trajectories highlighted in red, blue, orange, and purple.
    }
    \label{fig:Delta_E}
\end{figure}

Using the benchmark parameters in Tab.~\ref{table:refrence}, the energy variation as a function of the deformation rate $\Delta R_{\text{polar}}/R$ is obtained by numerically evaluating Eqs.~\eqref{eq:Delta_E1} and~\eqref{eq:Delta_E2}. The results, shown in the right panel of Fig.~\ref{fig:ball_Gamma_R}, are represented by two curves: the left pink curve corresponds to the prolate spheroidal bubble, and the right blue curve to the oblate spheroidal bubble, illustrating the energy differences for the two deformation modes.
For more general deformations, Fig.~\ref{fig:Delta_E} shows the three-dimensional energy difference landscape as a function of the deformation parameters \((b, c)\), with representative cases highlighted by the red, blue, orange, and purple curves.

As evident from Fig.~\ref{fig:ball_Gamma_R}, the energy variation reaches a minimum when the deformation vanishes, indicating that the bubble remains approximately spherical. This conclusion is further supported by the three-dimensional energy surface in Fig.~\ref{fig:Delta_E}, where the undeformed configuration ($b=c=R$) corresponds to a saddle point in energy space, suggesting stability against small tidal perturbations.

\subsection{Bubble nucleation rate near the black hole}
\label{subsec:bubblenearpbh}

We now turn to the case of bubble nucleation in the near-field region of the black hole, characterized by intense gravity and large spacetime curvature.
We begin this subsection by introducing a key assumption necessary for the validity of our theoretical framework:  
for any point in the vicinity of the black hole, the surface energy density $S_1$ associated with vacuum tunneling depends only on the local position.  
This assumption is motivated by the previous analyses of both bubbles nucleated at the black hole center and those formed far outside the horizon.  
In either case, the surface energy density of the tunneling configuration takes the universal form
\begin{equation}
    S_1 = k\sqrt{1 - \frac{2GM}{r}},
\end{equation}
where $r$ denotes the radial distance from the black hole center. It is worth noticing that this $r$ is different from $R$ in Eq.~\eqref{eq:S1} and $r_0$ in Eq.~\eqref{eq:S1_toy}. 
In the following, we adopt this expression for $S_1$ throughout our discussion.

\begin{figure}[h!]
    \centering
    \includegraphics[width=0.5\linewidth]{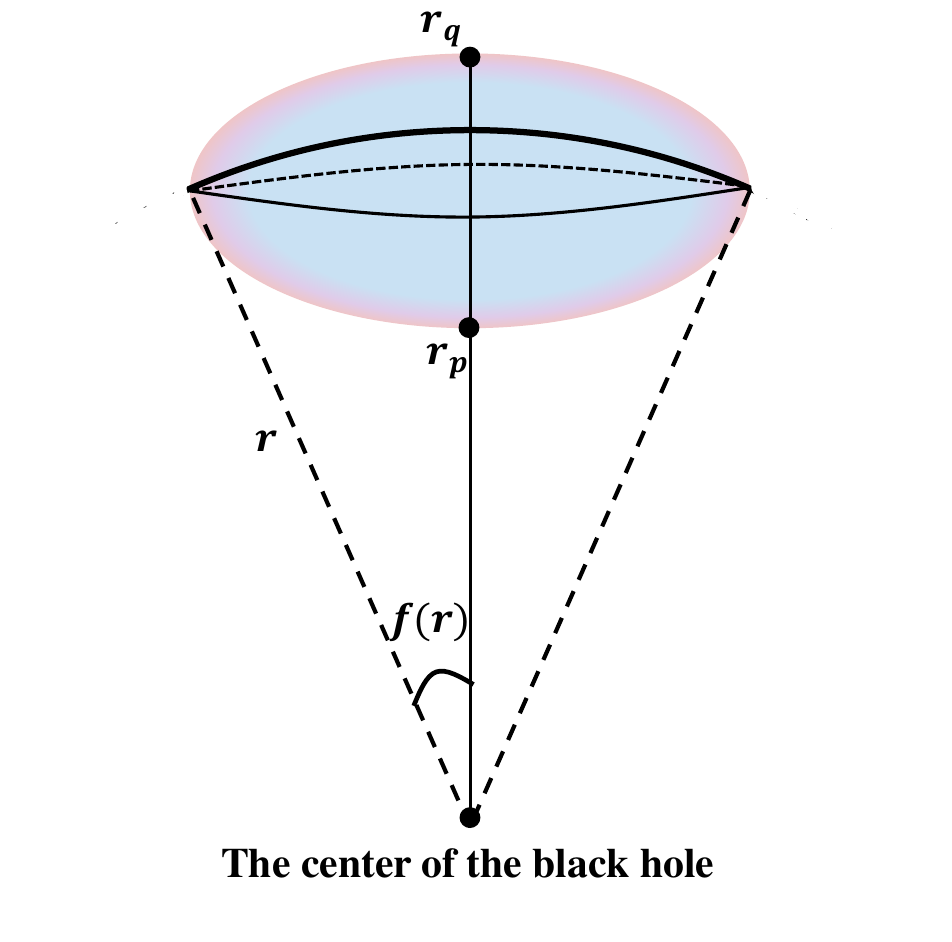}
    \caption{Illustration of a bubble constructed via spherical slicing in Schwarzschild spacetime.}
    \label{fig:illustration_of_f(r)}
\end{figure}

We consider a bubble whose nearest and farthest points from the black hole center are located at $r_p$ and $r_q$, respectively. Given this geometry and the spherical symmetry of Schwarzschild spacetime, we decompose the bubble into a series of concentric spherical shells centered on the black hole.  
For one spherical shell at radius $r$, we refer to its cross-section with the bubble as a \textit{spherical slice}, and the boundary of this slice as a \textit{boundary circle}.  Then the boundary of the bubble can be characterized by the polar angle of these circles as $\theta=f(r)$. The full geometric construction—including the parameters $r$, $r_p$, $r_q$, and the function $f(r)$—is illustrated in Fig.~\ref{fig:illustration_of_f(r)}.

Based on the analysis in Sec.~\ref{subsec:ballbubblevalidity}, we are justified in assuming that the bubble size remains much smaller than the Schwarzschild radius, so that each spherical cross-section can be properly approximated as locally planar.
This approximation allows the use of $f(r)$ to accurately describe the bubble contour without encountering multiple intersections or missed regions.

The bubble profile \( f(r) \) is determined by minimizing the Euclidean action. To this end, we begin by constructing the total action from the Wick-rotated Schwarzschild metric:
\begin{equation}
    S_4 = \int_0^{1/T} \sqrt{g_{00}}\, d\tau\, E_{\text{total}}
         = \int_0^{1/T} \sqrt{g_{00}}\, d\tau\, \left(-\epsilon V^{(3)}_{\text{bubble}} + S_1 S^{(2)}_{\text{bubble}}\right),
    \label{eq:S_4}
\end{equation}

where \(-\epsilon\) is the true vacuum energy density inside the bubble and \(S_1\) denotes the surface energy density of the bubble wall. To evaluate this action concretely, we now compute the geometric quantities \(V^{(3)}_{\text{bubble}}\) and \(S^{(2)}_{\text{bubble}}\).

The bubble volume can be obtained by integrating over spherical slices:
\begin{equation}
\begin{split}
    V^{(3)}_{\text{bubble}} &= \int_{r_p}^{r_q} dr\, \sqrt{g_{rr}}
         \int_0^{f(r)} d\theta
         \int_0^{2\pi} d\varphi\,
         \sqrt{g_{\theta\theta} g_{\varphi\varphi}} \\
      &= \int_{r_p}^{r_q} dr\, \sqrt{g_{rr}}\,
         2\pi r^2 [1 - \cos f(r)].
    \label{eq:V_ball}
\end{split}
\end{equation}

Similarly, the bubble surface area can be evaluated by integrating along the boundary circles:
\begin{equation}
    \begin{split}
        S^{(2)}_{\text{bubble}} &= \int_{L} dl \int_0^{2\pi} d\varphi \sqrt{g_{\varphi\varphi}} \\
          &= \int_{L} \sqrt{g_{\theta\theta}(d\theta)^2 + g_{rr}(dr)^2}\, 2\pi r\sin f(r) \\
          &= \int_{r_p}^{r_q} dr\, 2\pi r\sin f(r)
             \sqrt{g_{\theta\theta} (f')^2 + g_{rr}},
        \label{eq:S_ball}
    \end{split}
\end{equation}
where $g_{\varphi\varphi} = r^2\sin^2 f(r)$ and $f'(r) = d\theta/dr$.

Substituting Eqs.~\eqref{eq:V_ball} and~\eqref{eq:S_ball} into Eq.~\eqref{eq:S_4}, we obtain
\begin{equation}
    S_4 = \frac{1}{T} \int_{r_p}^{r_q} dr
          \left\{
              -2\pi\epsilon r^2 [1 - \cos f(r)]
              + 2\pi k r A \sin f(r)
                \sqrt{1 + r^2 A^2 (f')^2}
          \right\},
\end{equation}
where $A \equiv \sqrt{g_{00}} = \sqrt{1 - 2GM/r}$.

\vspace{0.5em}
Since the critical bubble size is much smaller than its distance from the black hole, we can safely assume $f(r) \ll 1$.  
Using the small-angle approximations $\sin f \approx f$ and $\cos f \approx 1 - f^2/2$, the action simplifies to
\begin{equation}
    S_4 = \frac{1}{T} \int_{r_p}^{r_q} dr
          \left[
              -\epsilon\pi r^2 f^2(r)
              + 2\pi k r A f(r)
                \sqrt{1 + r^2 A^2 (f')^2}
          \right].
    \label{S_4total}
\end{equation}

Applying the variational principle $\delta S_4 / \delta f = 0$ yields the Euler–Lagrange equation for $f(r)$:
\begin{equation}
    -\epsilon r^2 f
    + k r A \sqrt{1 + A^2 r^2 (f')^2}
    - \frac{d}{dr}
      \left[
          \frac{k r^3 A^3 f f'}{\sqrt{1 + A^2 r^2 (f')^2}}
      \right] = 0.
    \label{eq:finaleq}
\end{equation}

This equation cannot, in general, be solved analytically, and it often poses significant challenges for direct numerical integration. Nevertheless, by examining the asymptotic behavior of the bubble profile at the boundaries, where the first derivative diverges ($f'(r) \to \infty$ as $r \to r_p, r_q$), one can derive an approximate analytic expression for $f(r)$ in the vicinity of the boundary. Such boundary approximations can, in turn, substantially reduce the difficulty of numerically solving the full equation.

\vspace{0.5em}
In the vicinity of $r_p$, the derivative satisfies $f'(r) \big|_{r\to r_p} \to +\infty$.  
Expanding the relevant terms in powers of $(f')^{-2}$ gives
\begin{equation}
    \begin{split}
        \sqrt{1 + A^2 r^2 (f')^2}
        &\approx A r f'
           \left[ 1 + \frac{1}{2A^2 r^2 (f')^2}
           + \mathcal{O}(f'^{-4}) \right], \\
        \frac{k r^3 A^3 f f'}{\sqrt{1 + A^2 r^2 (f')^2}}
        &\approx k r^2 A^2 f
           \left[ 1 - \frac{1}{2A^2 r^2 (f')^2}
           + \mathcal{O}(f'^{-4}) \right].
    \end{split}
\end{equation}

Taking the derivative and keeping leading terms yields
\begin{equation}
    \frac{d}{dr}
    \left[
        \frac{k r^3 A^3 f f'}{\sqrt{1 + A^2 r^2 (f')^2}}
    \right]
    \approx 2k r A^2 f + 2k GM f + k r^2 A^2 f' - \frac{k}{2f'} + \mathcal{O}(f'^{-2}).
\end{equation}

Substituting this expansion into Eq.~(\ref{eq:finaleq}), one can simplify the equation to
\begin{equation}
    -\epsilon r^2 f - 2k r A^2 f - 2k GM f + \frac{k}{f'} = 0.
\end{equation}

This can be integrated to yield the near-boundary solution
\begin{equation}
    f_p^2(r) = \int_{r_p}^{r} dx\, 
    \frac{2k}{\epsilon x^2 + 2k x A^2 + 2k GM}.
    \label{eq:approx_f_p}
\end{equation}

\begin{figure}[]
    \centering
    \includegraphics[width=0.8\textwidth]{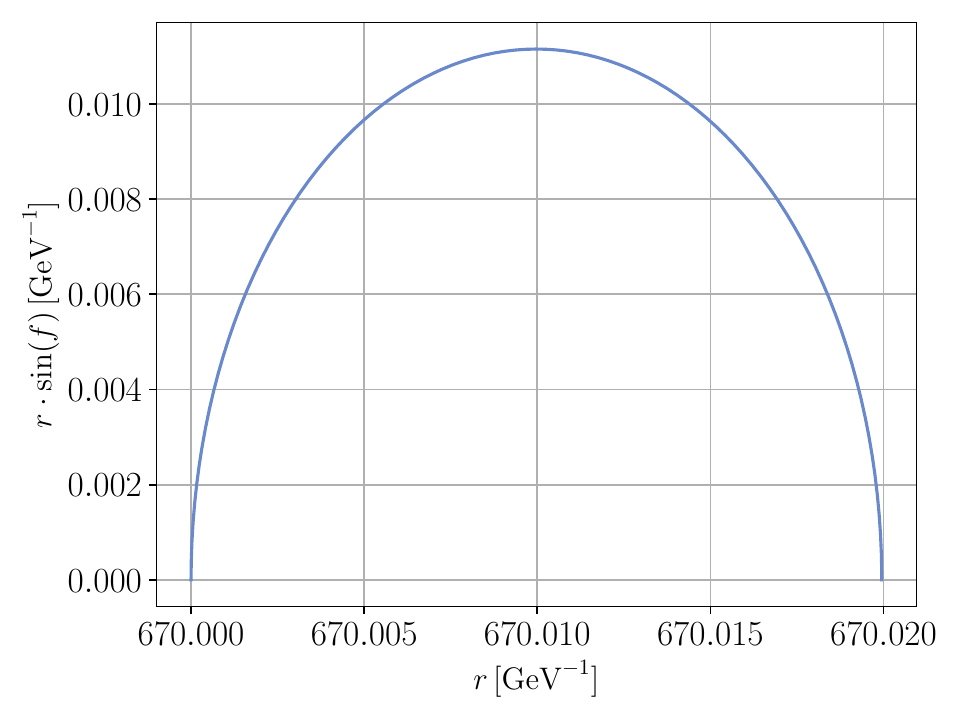}
    \caption{
    Numerical solution for $f(r)$ at $r_p = 10~GM$.  
    The bubble profile is obtained by plotting the boundary circle radius $r \sin f(r)$ as a function of $r$, and rotating the resulting curve around the horizontal axis to form the complete three-dimensional bubble surface.}
    \label{fig:profiel}
\end{figure}

By numerically solving Eq.~\eqref{eq:finaleq} using the approximate boundary condition from Eq.~\eqref{eq:approx_f_p}, we obtain the bubble profile as a function of distance from the black hole. The resulting contour, presented in Fig.~\ref{fig:profiel}, reveals that bubbles nucleated near the Schwarzschild black hole exhibit an oblate spheroidal shape. This morphology originates from the gravitational reduction of the bubble wall's surface energy density, which in turn lowers the compensating volume energy, thereby compressing the bubble. The compression is anisotropic, with the radial direction experiencing stronger deformation than the transverse direction due to its closer proximity to the black hole, ultimately producing the characteristic oblate form.

\begin{figure}[h]
    \centering
    \includegraphics[width=1.0\linewidth]{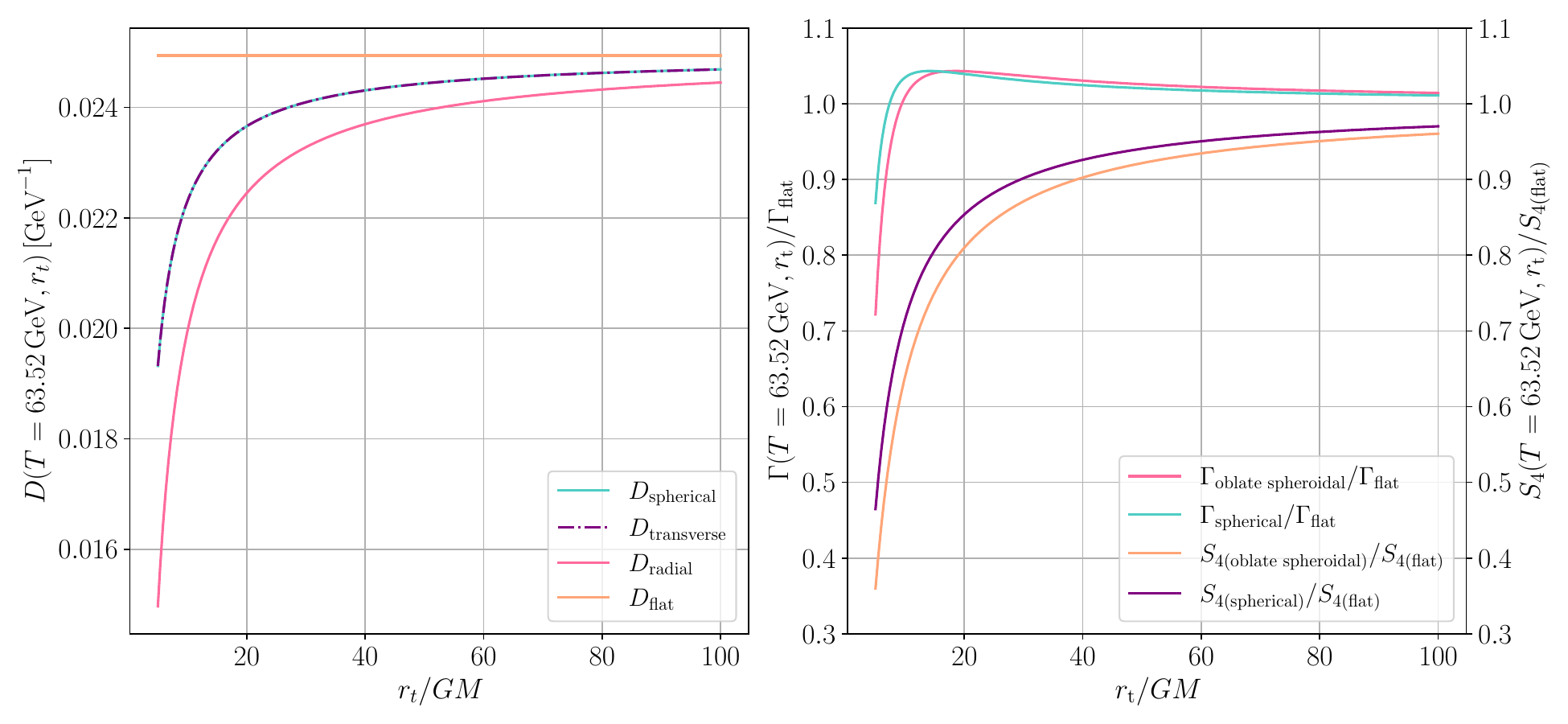}
    \caption{Left panel: Bubble diameters as functions of the distance $r_t$ from the the geometric center of the bubble to the black hole center.
    Right panel: Bubble nucleation rate and  Euclidean action, both normalized to their flat-space values, as functions of $r_t$.
    In the left panel, the orange line, $D_{\text{flat}}$, represents a bubble in Euclidean flat spacetime and remains constant, serving as a reference.  
    The bright cyan line, $D_{\text{spherical}}$, corresponds to a spherical bubble nucleated far from the black hole.  
    The pink line, $D_{\text{radial}}$, and the purple dash-dot line, $D_{\text{transverse}}$, denote the radial and transverse diameters of bubbles nucleated near the black hole. 
    In the right panel, the orange curve, $S_{4\text{(oblate spheroidal)}}/{S_{4\text{(flat)}}}$, shows the ratio of the Euclidean action for a bubble nucleated near the black hole to its flat-space value, while the purple curve, $S_{4\text{(spherical)}}/{S_{4\text{(flat)}}}$, denotes the same ratio for a spherical bubble. The pink curve, $\Gamma_{\text{oblate spheroidal}}/{\Gamma_\text{flat}}$, and the bright-cyan curve, $\Gamma_{\text{spherical}}/{\Gamma_\text{flat}}$, represent the corresponding nucleation-rate ratios for oblate and spherical bubbles, respectively. 
    }
    \label{D-rGamma-r}
\end{figure}

 Using Eqs.~\eqref{S_4total} and~\eqref{concise_form_of_Gamma}, we further compute the critical bubble diameters, Euclidean actions, and nucleation rates, all shown in Fig.~\ref{D-rGamma-r} as functions of the distance $r_t$ from the bubble's geometric center to the black hole center.
 The left panel displays the radial and transverse diameters, represented by the pink line and the purple dash-dot line, respectively. Both diameters decrease at small $r_t$ relative to the flat-spacetime reference shown as the orange line, indicating significant gravitational compression. The transverse diameter remains smaller than the radial diameter, confirming the oblate shape, with the difference diminishing as $r_t$ increases. All curves converge to the flat-spacetime value at large $r_t$. In the right panel, the orange curve shows the normalized action $S_{4\text{(oblate spheroidal)}}/{S_{4\text{(flat)}}}$ for the oblate spheroidal bubble, which decreases with decreasing $r_t$ and approaches unity in the weak-field limit. The pink curve gives the corresponding normalized nucleation rate $\Gamma_{\text{oblate spheroidal}}/{\Gamma_{\text{flat}}}$, which increases from below unity at small $r_t$ to above unity at moderate $r_t$, before converging to unity in the weak-field regime. These results show that black-hole gravity reduces the Euclidean action, thereby enhancing the nucleation rate in moderate fields, while both quantities recover their flat-space values when gravitational effects become negligible. At very small $r_t$, however, the nucleation rate is suppressed relative to the flat-spacetime case even though the action is significantly reduced, which may indicate that Eq.~\eqref{concise_form_of_Gamma} becomes less reliable in the strong-gravity regime.

When compared with the spherical case, the oblate bubble exhibits qualitatively similar trends but experiences a stronger gravitational suppression. This comparison allows us to estimate the effective regime for the spherical approximation as approximately \( r_t \gtrsim 80~GM \), as evidenced by the close alignment between the radial diameter and the spherical solution in this region. Furthermore, the coincidence of \( D_{\text{transverse}} \) and \( D_{\text{spherical}} \) at large \( r_t \) provides another notable feature, which may be explained by \( r_t \) corresponding to the location of the maximal transverse diameter, where the critical cross-sectional radius matches that of the spherical bubble solution.

\section{Discussions on new gravitational sources from deformed bubbles}
\label{sec:discussion of gw}

Based on the analysis in previous sections, we have shown that a PBH induces deformation of spherical bubbles when nucleation occurs away from the black hole center. This deformation of the bubble shape has direct implications for the gravitational wave signals generated during bubble expansion and collision. For perfectly spherical bubbles, the system possesses exact spherical symmetry, resulting in a vanishing quadrupole moment; hence, gravitational waves can be produced through bubble collisions when the spherical symmetry is broken. In contrast, the oblate spheroidal bubbles studied in this work are inherently non-spherical, leading to a nonvanishing second time derivative of the quadrupole moment even during expansion. This can be demonstrated through a simple illustrative model.

Consider an axisymmetric oblate spheroid that rotates about its short axis and undergoes a self-similar expansion. Its semi-major and semi-minor axes are parameterized by \( \mathcal{A}(t) = v t \) and \( \mathcal{B}(t) = \mathcal{B}_0 v t \), respectively, where $v$ is the bubble expansion velocity and \( \mathcal{B}_0 < 1 \) is the fixed aspect ratio. The surface of this ellipsoid is then given by:
\begin{equation}
    \frac{x^2}{\mathcal{A}^2} + \frac{y^2}{\mathcal{A}^2} + \frac{z^2}{\mathcal{B}^2} = 1.
\end{equation}

If the energy density is uniform ($\rho = \text{const}$), the third time derivative of the quadrupole tensor of the ellipsoid is given by
\begin{equation}
\begin{split}
    \dddot{Q}_{ij} &= \partial_t^3 \int \rho \left(3r_i r_j - r^2 \delta_{ij}\right) dV \\
    &= \partial_t^3 \left[ \frac{4\pi\rho}{15}(\mathcal{A}^4 \mathcal{B} - \mathcal{A}^2 \mathcal{B}^3)
    \begin{pmatrix}
        1 & & \\
        & 1 & \\
        & & -2
    \end{pmatrix} \right] \\
    &= 60 v^5 t^2 (\mathcal{B}_0 - \mathcal{B}_0^3)
    \begin{pmatrix}
        1 & & \\
        & 1 & \\
        & & -2
    \end{pmatrix} \neq 0.
\end{split}
\end{equation}

This nonvanishing time-dependent result demonstrates that gravitational waves can be emitted even during the expansion of a single non-spherical bubble, as the power radiated obeys \( P_{\mathrm{GW}} \propto G|\dddot{Q}_{ij}|^2 \).
In the presence of a black hole background, the anisotropic gravitational field can further modify the expansion velocities along different directions of the bubble, thereby enhancing the asymmetry and potentially amplifying the gravitational wave emission. This simple estimation indicates a new source of phase transition gravitational wave before the bubbles collide. Detailed study on this new gravitational wave source is left in our future work.

\section{Conclusions}\label{sec:sum}

In this work, we have investigated in detail how the presence of PBHs affects the bubble profile, the corresponding Euclidean action and nucleation rate during a cosmological FOPT. Our analysis shows that, whether the nucleation occurs near the horizon of a black hole—where the black hole serves as a nucleation seed—or outside its vicinity, the resulting bubbles generally possess smaller radii compared to those nucleated in a flat spacetime background. Consequently, the corresponding bounce action is reduced, which may leads to a modification of the overall bubble nucleation rate.

For the case in which a bubble nucleates with the black hole at its center, the method employed in this work is valid only when the bubble radius is much larger than the black hole horizon scale. Numerical estimates show that this requirement corresponds to a black hole mass smaller than $10^{34}\,\mathrm{GeV}$, for which the ratio between the bubble radius and the Schwarzschild radius is approximately $R_{\text{bubble}}/R_{\text{BH}} \simeq 92.8$. In this regime, the presence of a PBH reduces the bubble radius, thereby lowering the corresponding Euclidean action and enhancing the nucleation rate.

For bubbles nucleating at distances far from the black hole, namely for $r_t \gtrsim 80~GM$, the local spacetime can be treated as approximately asymptotically flat. Under this assumption, our calculations indicate that the nucleated bubbles are stable under small perturbations, which further supports the validity of our conclusions obtained in flat spacetime. In such configurations, the PBH leads to a suppression of the bubble radius and a corresponding decrease in the Euclidean action. This promotes the nucleation rate in moderate-to-weak gravitational environments, while in strong-field regions the rate might become suppressed—an effect that may stem from the breakdown of the nucleation rate formula under extreme gravitational conditions.

For the case where bubbles nucleate in the vicinity of a black hole, we utilized the intrinsic $\mathrm{O}(1)$ symmetry of the system to construct the corresponding action, and numerically solved the bubble profile from first principles. We observed that under strong gravitational fields, the bubble is compressed and deforms into an oblate spheroidal shape, which leads to a reduction in the Euclidean action and modifies the nucleation rate in a manner consistent with the spherical case. As the distance between the bubble and the black hole increases, the gravitational distortion gradually weakens, and the bubble shape asymptotically approaches a spherical configuration similar to that in flat spacetime.

These results indicate that PBHs can significantly influence the dynamics of cosmological phase transitions by catalyzing bubble formation and modifying the effective nucleation probability. These findings suggest that PBHs may play a nontrivial role in catalyzing vacuum decay and shaping the gravitational wave spectrum originating from FOPTs in the early Universe. Such an effect may have important implications for understanding the interplay between PBH populations, early-Universe phase transition dynamics, the vacuum decay process, the multi-messenger signatures of early-Universe cosmology, and its potential gravitational wave signatures.
Our analysis of the influence of  inhomogeneous sources such as PBHs on false vacuum decay and bubble nucleation rates may also be applicable to other possible inhomogeneous gravitational sources in the early universe.


\begin{acknowledgments} 
This work is supported by the National Natural Science Foundation of China (NNSFC) Grant No.12475111, No. 12205387,
and the Undergraduate Research and Innovation Project.

\end{acknowledgments}


\bibliography{PBH_refs}

\begin{thebibliography}{58}%
\makeatletter
\providecommand \@ifxundefined [1]{%
 \@ifx{#1\undefined}
}%
\providecommand \@ifnum [1]{%
 \ifnum #1\expandafter \@firstoftwo
 \else \expandafter \@secondoftwo
 \fi
}%
\providecommand \@ifx [1]{%
 \ifx #1\expandafter \@firstoftwo
 \else \expandafter \@secondoftwo
 \fi
}%
\providecommand \natexlab [1]{#1}%
\providecommand \enquote  [1]{``#1''}%
\providecommand \bibnamefont  [1]{#1}%
\providecommand \bibfnamefont [1]{#1}%
\providecommand \citenamefont [1]{#1}%
\providecommand \href@noop [0]{\@secondoftwo}%
\providecommand \href [0]{\begingroup \@sanitize@url \@href}%
\providecommand \@href[1]{\@@startlink{#1}\@@href}%
\providecommand \@@href[1]{\endgroup#1\@@endlink}%
\providecommand \@sanitize@url [0]{\catcode `\\12\catcode `\$12\catcode `\&12\catcode `\#12\catcode `\^12\catcode `\_12\catcode `\%12\relax}%
\providecommand \@@startlink[1]{}%
\providecommand \@@endlink[0]{}%
\providecommand \url  [0]{\begingroup\@sanitize@url \@url }%
\providecommand \@url [1]{\endgroup\@href {#1}{\urlprefix }}%
\providecommand \urlprefix  [0]{URL }%
\providecommand \Eprint [0]{\href }%
\providecommand \doibase [0]{http://dx.doi.org/}%
\providecommand \selectlanguage [0]{\@gobble}%
\providecommand \bibinfo  [0]{\@secondoftwo}%
\providecommand \bibfield  [0]{\@secondoftwo}%
\providecommand \translation [1]{[#1]}%
\providecommand \BibitemOpen [0]{}%
\providecommand \bibitemStop [0]{}%
\providecommand \bibitemNoStop [0]{.\EOS\space}%
\providecommand \EOS [0]{\spacefactor3000\relax}%
\providecommand \BibitemShut  [1]{\csname bibitem#1\endcsname}%
\let\auto@bib@innerbib\@empty
\bibitem [{\citenamefont {Huang}(2025)}]{Huang:2025cqi}%
  \BibitemOpen
  \bibfield  {author} {\bibinfo {author} {\bibfnamefont {F.~P.}\ \bibnamefont {Huang}},\ }\href@noop {} {\  (\bibinfo {year} {2025})},\ \Eprint {http://arxiv.org/abs/2501.15543} {arXiv:2501.15543 [hep-ph]} \BibitemShut {NoStop}%
\bibitem [{\citenamefont {White}(2022)}]{White:2022ufa}%
  \BibitemOpen
  \bibfield  {author} {\bibinfo {author} {\bibfnamefont {G.}~\bibnamefont {White}},\ }\href {\doibase 10.1088/978-0-7503-3571-3} {\emph {\bibinfo {title} {{Electroweak Baryogenesis (Second Edition)}}}}\ (\bibinfo  {publisher} {IOP},\ \bibinfo {year} {2022})\BibitemShut {NoStop}%
\bibitem [{\citenamefont {van~de Vis}\ \emph {et~al.}(2025)\citenamefont {van~de Vis}, \citenamefont {de~Vries},\ and\ \citenamefont {Postma}}]{vandeVis:2025efm}%
  \BibitemOpen
  \bibfield  {author} {\bibinfo {author} {\bibfnamefont {J.}~\bibnamefont {van~de Vis}}, \bibinfo {author} {\bibfnamefont {J.}~\bibnamefont {de~Vries}}, \ and\ \bibinfo {author} {\bibfnamefont {M.}~\bibnamefont {Postma}},\ }\href@noop {} {\  (\bibinfo {year} {2025})},\ \Eprint {http://arxiv.org/abs/2508.09989} {arXiv:2508.09989 [hep-ph]} \BibitemShut {NoStop}%
\bibitem [{\citenamefont {Morrissey}\ and\ \citenamefont {Ramsey-Musolf}(2012)}]{Morrissey:2012db}%
  \BibitemOpen
  \bibfield  {author} {\bibinfo {author} {\bibfnamefont {D.~E.}\ \bibnamefont {Morrissey}}\ and\ \bibinfo {author} {\bibfnamefont {M.~J.}\ \bibnamefont {Ramsey-Musolf}},\ }\href {\doibase 10.1088/1367-2630/14/12/125003} {\bibfield  {journal} {\bibinfo  {journal} {New J. Phys.}\ }\textbf {\bibinfo {volume} {14}},\ \bibinfo {pages} {125003} (\bibinfo {year} {2012})},\ \Eprint {http://arxiv.org/abs/1206.2942} {arXiv:1206.2942 [hep-ph]} \BibitemShut {NoStop}%
\bibitem [{\citenamefont {Witten}(1984)}]{witten:1984}%
  \BibitemOpen
  \bibfield  {author} {\bibinfo {author} {\bibfnamefont {E.}~\bibnamefont {Witten}},\ }\href {\doibase 10.1103/PhysRevD.30.272} {\bibfield  {journal} {\bibinfo  {journal} {Phys. Rev. D}\ }\textbf {\bibinfo {volume} {30}},\ \bibinfo {pages} {272} (\bibinfo {year} {1984})}\BibitemShut {NoStop}%
\bibitem [{\citenamefont {Krylov}\ \emph {et~al.}(2013)\citenamefont {Krylov}, \citenamefont {Levin},\ and\ \citenamefont {Rubakov}}]{Krylov:2013qe}%
  \BibitemOpen
  \bibfield  {author} {\bibinfo {author} {\bibfnamefont {E.}~\bibnamefont {Krylov}}, \bibinfo {author} {\bibfnamefont {A.}~\bibnamefont {Levin}}, \ and\ \bibinfo {author} {\bibfnamefont {V.}~\bibnamefont {Rubakov}},\ }\href {\doibase 10.1103/PhysRevD.87.083528} {\bibfield  {journal} {\bibinfo  {journal} {Phys. Rev. D}\ }\textbf {\bibinfo {volume} {87}},\ \bibinfo {pages} {083528} (\bibinfo {year} {2013})},\ \Eprint {http://arxiv.org/abs/1301.0354} {arXiv:1301.0354 [hep-ph]} \BibitemShut {NoStop}%
\bibitem [{\citenamefont {Huang}\ and\ \citenamefont {Li}(2017)}]{Huang:2017kzu}%
  \BibitemOpen
  \bibfield  {author} {\bibinfo {author} {\bibfnamefont {F.~P.}\ \bibnamefont {Huang}}\ and\ \bibinfo {author} {\bibfnamefont {C.~S.}\ \bibnamefont {Li}},\ }\href {\doibase 10.1103/PhysRevD.96.095028} {\bibfield  {journal} {\bibinfo  {journal} {Phys. Rev. D}\ }\textbf {\bibinfo {volume} {96}},\ \bibinfo {pages} {095028} (\bibinfo {year} {2017})},\ \Eprint {http://arxiv.org/abs/1709.09691} {arXiv:1709.09691 [hep-ph]} \BibitemShut {NoStop}%
\bibitem [{\citenamefont {Jiang}\ \emph {et~al.}(2024{\natexlab{a}})\citenamefont {Jiang}, \citenamefont {Yang}, \citenamefont {Ma},\ and\ \citenamefont {Huang}}]{Jiang:2023qbm}%
  \BibitemOpen
  \bibfield  {author} {\bibinfo {author} {\bibfnamefont {S.}~\bibnamefont {Jiang}}, \bibinfo {author} {\bibfnamefont {A.}~\bibnamefont {Yang}}, \bibinfo {author} {\bibfnamefont {J.}~\bibnamefont {Ma}}, \ and\ \bibinfo {author} {\bibfnamefont {F.~P.}\ \bibnamefont {Huang}},\ }\href {\doibase 10.1088/1361-6382/ad24c6} {\bibfield  {journal} {\bibinfo  {journal} {Class. Quant. Grav.}\ }\textbf {\bibinfo {volume} {41}},\ \bibinfo {pages} {065009} (\bibinfo {year} {2024}{\natexlab{a}})},\ \Eprint {http://arxiv.org/abs/2306.17827} {arXiv:2306.17827 [hep-ph]} \BibitemShut {NoStop}%
\bibitem [{\citenamefont {Baker}\ \emph {et~al.}(2020)\citenamefont {Baker}, \citenamefont {Kopp},\ and\ \citenamefont {Long}}]{Baker:2020}%
  \BibitemOpen
  \bibfield  {author} {\bibinfo {author} {\bibfnamefont {M.~J.}\ \bibnamefont {Baker}}, \bibinfo {author} {\bibfnamefont {J.}~\bibnamefont {Kopp}}, \ and\ \bibinfo {author} {\bibfnamefont {A.~J.}\ \bibnamefont {Long}},\ }\href {\doibase 10.1103/PhysRevLett.125.151102} {\bibfield  {journal} {\bibinfo  {journal} {Phys. Rev. Lett.}\ }\textbf {\bibinfo {volume} {125}},\ \bibinfo {pages} {151102} (\bibinfo {year} {2020})}\BibitemShut {NoStop}%
\bibitem [{\citenamefont {Chway}\ \emph {et~al.}(2020)\citenamefont {Chway}, \citenamefont {Jung},\ and\ \citenamefont {Shin}}]{Chway:2019kft}%
  \BibitemOpen
  \bibfield  {author} {\bibinfo {author} {\bibfnamefont {D.}~\bibnamefont {Chway}}, \bibinfo {author} {\bibfnamefont {T.~H.}\ \bibnamefont {Jung}}, \ and\ \bibinfo {author} {\bibfnamefont {C.~S.}\ \bibnamefont {Shin}},\ }\href {\doibase 10.1103/PhysRevD.101.095019} {\bibfield  {journal} {\bibinfo  {journal} {Phys. Rev. D}\ }\textbf {\bibinfo {volume} {101}},\ \bibinfo {pages} {095019} (\bibinfo {year} {2020})},\ \Eprint {http://arxiv.org/abs/1912.04238} {arXiv:1912.04238 [hep-ph]} \BibitemShut {NoStop}%
\bibitem [{\citenamefont {Azatov}\ \emph {et~al.}(2021)\citenamefont {Azatov}, \citenamefont {Vanvlasselaer},\ and\ \citenamefont {Yin}}]{Azatov:2021ifm}%
  \BibitemOpen
  \bibfield  {author} {\bibinfo {author} {\bibfnamefont {A.}~\bibnamefont {Azatov}}, \bibinfo {author} {\bibfnamefont {M.}~\bibnamefont {Vanvlasselaer}}, \ and\ \bibinfo {author} {\bibfnamefont {W.}~\bibnamefont {Yin}},\ }\href {\doibase 10.1007/JHEP03(2021)288} {\bibfield  {journal} {\bibinfo  {journal} {JHEP}\ }\textbf {\bibinfo {volume} {03}},\ \bibinfo {pages} {288} (\bibinfo {year} {2021})},\ \Eprint {http://arxiv.org/abs/2101.05721} {arXiv:2101.05721 [hep-ph]} \BibitemShut {NoStop}%
\bibitem [{\citenamefont {Jiang}\ \emph {et~al.}(2024{\natexlab{b}})\citenamefont {Jiang}, \citenamefont {Huang},\ and\ \citenamefont {Ko}}]{Jiang:2024zrb}%
  \BibitemOpen
  \bibfield  {author} {\bibinfo {author} {\bibfnamefont {S.}~\bibnamefont {Jiang}}, \bibinfo {author} {\bibfnamefont {F.~P.}\ \bibnamefont {Huang}}, \ and\ \bibinfo {author} {\bibfnamefont {P.}~\bibnamefont {Ko}},\ }\href {\doibase 10.1007/JHEP07(2024)053} {\bibfield  {journal} {\bibinfo  {journal} {JHEP}\ }\textbf {\bibinfo {volume} {07}},\ \bibinfo {pages} {053} (\bibinfo {year} {2024}{\natexlab{b}})},\ \Eprint {http://arxiv.org/abs/2404.16509} {arXiv:2404.16509 [hep-ph]} \BibitemShut {NoStop}%
\bibitem [{com(2025)}]{combinedPBHRefs}%
  \BibitemOpen
  \href@noop {} {\enquote {\bibinfo {title} {{S. W. Hawking et al.}, \href{https://doi.org/10.1103/PhysRevD.26.2681}{Phys. Rev. D 26, 2681 (1982)}; {H. Kodama et al.}, \href{https://doi.org/10.1143/PTP.68.1979}{Prog. Theor. Phys. 68, 1979 (1982)}; {I. G. Moss}, \href{https://arxiv.org/abs/gr-qc/9405045}{arXiv:gr-qc/9405045 (1994)}; {R. V. Konoplich et al.}, \href{https://doi.org/10.1134/1.1359486}{Phys. Atom. Nucl. 62, 1593 (1999)}; {H. Deng \& A. Vilenkin}, \href{https://doi.org/10.1088/1475-7516/2017/12/044}{JCAP 12, 044 (2017)}; {H. Deng}, \href{https://doi.org/10.1088/1475-7516/2020/09/023}{JCAP 09, 023 (2020)}; {K. Kawana \& K.-P. Xie}, \href{https://doi.org/10.1016/j.physletb.2021.136791}{Phys. Lett. B 824, 136791 (2022)}; {K. Hashino et al.}, \href{https://doi.org/10.1016/j.physletb.2022.137261}{Phys. Lett. B 833, 137261 (2022)}; {Y. Gouttenoire \& T. Volansky}, \href{https://doi.org/10.1103/PhysRevD.110.043514}{Phys. Rev. D 110, 043514 (2024)}; {Y. Gouttenoire},
  \href{https://doi.org/10.1016/j.physletb.2024.138800}{Phys. Lett. B 855, 138800 (2024)}; {S. Kanemura et al.}, \href{https://doi.org/10.1007/JHEP06(2024)036}{JHEP 06, 036 (2024)}; d{. Gonçalves et al.}, \href{https://doi.org/10.1103/PhysRevD.111.035009}{Phys. Rev. D 111, 035009 (2025)}; {S. Balaji et al.}, \href{https://doi.org/10.1103/PhysRevD.109.075048}{Phys. Rev. D 109, 075048 (2024)}; {M. Lewicki et al.}, \href{https://doi.org/10.1016/j.dark.2025.102075}{Phys. Dark Univ. 50, 102075 (2025)}; {K. Hashino et al.}, \href{https://doi.org/10.1088/1475-7516/2025/09/006}{JCAP 09, 006 (2025)}; {K. Murai et al.}, \href{https://doi.org/10.1007/JHEP07(2025)065}{JHEP 07, 065 (2025)}; {F. P. Huang et al.}, \href{https://arxiv.org/abs/2510.24007}{arXiv:2510.24007 (2025)}; {J. Liu et al.}, \href{https://doi.org/10.1103/PhysRevD.105.L021303}{Phys. Rev. D 105, L021303 (2022)}},}\ } (\bibinfo {year} {1982--2025})\BibitemShut {NoStop}%
\bibitem [{\citenamefont {Vachaspati}(1991)}]{Vachaspati:1991nm}%
  \BibitemOpen
  \bibfield  {author} {\bibinfo {author} {\bibfnamefont {T.}~\bibnamefont {Vachaspati}},\ }\href {\doibase 10.1016/0370-2693(91)90051-Q} {\bibfield  {journal} {\bibinfo  {journal} {Phys. Lett. B}\ }\textbf {\bibinfo {volume} {265}},\ \bibinfo {pages} {258} (\bibinfo {year} {1991})}\BibitemShut {NoStop}%
\bibitem [{\citenamefont {Huber}\ and\ \citenamefont {Konstandin}(2008)}]{Huber:2008hg}%
  \BibitemOpen
  \bibfield  {author} {\bibinfo {author} {\bibfnamefont {S.~J.}\ \bibnamefont {Huber}}\ and\ \bibinfo {author} {\bibfnamefont {T.}~\bibnamefont {Konstandin}},\ }\href {\doibase 10.1088/1475-7516/2008/09/022} {\bibfield  {journal} {\bibinfo  {journal} {JCAP}\ }\textbf {\bibinfo {volume} {09}},\ \bibinfo {pages} {022} (\bibinfo {year} {2008})},\ \Eprint {http://arxiv.org/abs/0806.1828} {arXiv:0806.1828 [hep-ph]} \BibitemShut {NoStop}%
\bibitem [{\citenamefont {Caprini}\ \emph {et~al.}(2016)\citenamefont {Caprini} \emph {et~al.}}]{Caprini:2015zlo}%
  \BibitemOpen
  \bibfield  {author} {\bibinfo {author} {\bibfnamefont {C.}~\bibnamefont {Caprini}} \emph {et~al.},\ }\href {\doibase 10.1088/1475-7516/2016/04/001} {\bibfield  {journal} {\bibinfo  {journal} {JCAP}\ }\textbf {\bibinfo {volume} {04}},\ \bibinfo {pages} {001} (\bibinfo {year} {2016})},\ \Eprint {http://arxiv.org/abs/1512.06239} {arXiv:1512.06239 [astro-ph.CO]} \BibitemShut {NoStop}%
\bibitem [{\citenamefont {Hindmarsh}\ \emph {et~al.}(2017)\citenamefont {Hindmarsh}, \citenamefont {Huber}, \citenamefont {Rummukainen},\ and\ \citenamefont {Weir}}]{Hindmarsh:2017gnf}%
  \BibitemOpen
  \bibfield  {author} {\bibinfo {author} {\bibfnamefont {M.}~\bibnamefont {Hindmarsh}}, \bibinfo {author} {\bibfnamefont {S.~J.}\ \bibnamefont {Huber}}, \bibinfo {author} {\bibfnamefont {K.}~\bibnamefont {Rummukainen}}, \ and\ \bibinfo {author} {\bibfnamefont {D.~J.}\ \bibnamefont {Weir}},\ }\href {\doibase 10.1103/PhysRevD.96.103520} {\bibfield  {journal} {\bibinfo  {journal} {Phys. Rev. D}\ }\textbf {\bibinfo {volume} {96}},\ \bibinfo {pages} {103520} (\bibinfo {year} {2017})},\ \bibinfo {note} {[Erratum: Phys.Rev.D 101, 089902 (2020)]},\ \Eprint {http://arxiv.org/abs/1704.05871} {arXiv:1704.05871 [astro-ph.CO]} \BibitemShut {NoStop}%
\bibitem [{\citenamefont {Qiu}\ \emph {et~al.}(2025)\citenamefont {Qiu}, \citenamefont {Jiang},\ and\ \citenamefont {Huang}}]{Qiu:2025tmn}%
  \BibitemOpen
  \bibfield  {author} {\bibinfo {author} {\bibfnamefont {D.}~\bibnamefont {Qiu}}, \bibinfo {author} {\bibfnamefont {S.}~\bibnamefont {Jiang}}, \ and\ \bibinfo {author} {\bibfnamefont {F.~P.}\ \bibnamefont {Huang}},\ }\href@noop {} {\  (\bibinfo {year} {2025})},\ \Eprint {http://arxiv.org/abs/2508.04314} {arXiv:2508.04314 [hep-ph]} \BibitemShut {NoStop}%
\bibitem [{\citenamefont {Coleman}(1977)}]{PhysRevD.15.2929}%
  \BibitemOpen
  \bibfield  {author} {\bibinfo {author} {\bibfnamefont {S.}~\bibnamefont {Coleman}},\ }\href {\doibase 10.1103/PhysRevD.15.2929} {\bibfield  {journal} {\bibinfo  {journal} {Phys. Rev. D}\ }\textbf {\bibinfo {volume} {15}},\ \bibinfo {pages} {2929} (\bibinfo {year} {1977})}\BibitemShut {NoStop}%
\bibitem [{\citenamefont {Callan}\ and\ \citenamefont {Coleman}(1977)}]{PhysRevD.16.1762}%
  \BibitemOpen
  \bibfield  {author} {\bibinfo {author} {\bibfnamefont {C.~G.}\ \bibnamefont {Callan}}\ and\ \bibinfo {author} {\bibfnamefont {S.}~\bibnamefont {Coleman}},\ }\href {\doibase 10.1103/PhysRevD.16.1762} {\bibfield  {journal} {\bibinfo  {journal} {Phys. Rev. D}\ }\textbf {\bibinfo {volume} {16}},\ \bibinfo {pages} {1762} (\bibinfo {year} {1977})}\BibitemShut {NoStop}%
\bibitem [{\citenamefont {Jinno}\ and\ \citenamefont {Sato}(2021)}]{Jinno:2021}%
  \BibitemOpen
  \bibfield  {author} {\bibinfo {author} {\bibfnamefont {R.}~\bibnamefont {Jinno}}\ and\ \bibinfo {author} {\bibfnamefont {R.}~\bibnamefont {Sato}},\ }\href {\doibase 10.1103/PhysRevD.104.096009} {\bibfield  {journal} {\bibinfo  {journal} {Phys. Rev. D}\ }\textbf {\bibinfo {volume} {104}},\ \bibinfo {pages} {096009} (\bibinfo {year} {2021})}\BibitemShut {NoStop}%
\bibitem [{\citenamefont {Hayashi}\ \emph {et~al.}(2022)\citenamefont {Hayashi}, \citenamefont {Kamada}, \citenamefont {Oshita},\ and\ \citenamefont {Yokoyama}}]{Hayashi:2021kro}%
  \BibitemOpen
  \bibfield  {author} {\bibinfo {author} {\bibfnamefont {T.}~\bibnamefont {Hayashi}}, \bibinfo {author} {\bibfnamefont {K.}~\bibnamefont {Kamada}}, \bibinfo {author} {\bibfnamefont {N.}~\bibnamefont {Oshita}}, \ and\ \bibinfo {author} {\bibfnamefont {J.}~\bibnamefont {Yokoyama}},\ }\href {\doibase 10.1088/1475-7516/2022/05/041} {\bibfield  {journal} {\bibinfo  {journal} {JCAP}\ }\textbf {\bibinfo {volume} {05}},\ \bibinfo {pages} {041} (\bibinfo {year} {2022})},\ \Eprint {http://arxiv.org/abs/2112.09284} {arXiv:2112.09284 [hep-th]} \BibitemShut {NoStop}%
\bibitem [{\citenamefont {Ai}\ \emph {et~al.}(2019)\citenamefont {Ai}, \citenamefont {Garbrecht},\ and\ \citenamefont {Tamarit}}]{Ai:2019fri}%
  \BibitemOpen
  \bibfield  {author} {\bibinfo {author} {\bibfnamefont {W.-Y.}\ \bibnamefont {Ai}}, \bibinfo {author} {\bibfnamefont {B.}~\bibnamefont {Garbrecht}}, \ and\ \bibinfo {author} {\bibfnamefont {C.}~\bibnamefont {Tamarit}},\ }\href {\doibase 10.1007/JHEP12(2019)095} {\bibfield  {journal} {\bibinfo  {journal} {JHEP}\ }\textbf {\bibinfo {volume} {12}},\ \bibinfo {pages} {095} (\bibinfo {year} {2019})},\ \Eprint {http://arxiv.org/abs/1905.04236} {arXiv:1905.04236 [hep-th]} \BibitemShut {NoStop}%
\bibitem [{\citenamefont {Andreassen}\ \emph {et~al.}(2017)\citenamefont {Andreassen}, \citenamefont {Farhi}, \citenamefont {Frost},\ and\ \citenamefont {Schwartz}}]{Andreassen:2016cvx}%
  \BibitemOpen
  \bibfield  {author} {\bibinfo {author} {\bibfnamefont {A.}~\bibnamefont {Andreassen}}, \bibinfo {author} {\bibfnamefont {D.}~\bibnamefont {Farhi}}, \bibinfo {author} {\bibfnamefont {W.}~\bibnamefont {Frost}}, \ and\ \bibinfo {author} {\bibfnamefont {M.~D.}\ \bibnamefont {Schwartz}},\ }\href {\doibase 10.1103/PhysRevD.95.085011} {\bibfield  {journal} {\bibinfo  {journal} {Phys. Rev. D}\ }\textbf {\bibinfo {volume} {95}},\ \bibinfo {pages} {085011} (\bibinfo {year} {2017})},\ \Eprint {http://arxiv.org/abs/1604.06090} {arXiv:1604.06090 [hep-th]} \BibitemShut {NoStop}%
\bibitem [{\citenamefont {Andreassen}\ \emph {et~al.}(2016)\citenamefont {Andreassen}, \citenamefont {Farhi}, \citenamefont {Frost},\ and\ \citenamefont {Schwartz}}]{PhysRevLett.117.231601}%
  \BibitemOpen
  \bibfield  {author} {\bibinfo {author} {\bibfnamefont {A.}~\bibnamefont {Andreassen}}, \bibinfo {author} {\bibfnamefont {D.}~\bibnamefont {Farhi}}, \bibinfo {author} {\bibfnamefont {W.}~\bibnamefont {Frost}}, \ and\ \bibinfo {author} {\bibfnamefont {M.~D.}\ \bibnamefont {Schwartz}},\ }\href {\doibase 10.1103/PhysRevLett.117.231601} {\bibfield  {journal} {\bibinfo  {journal} {Phys. Rev. Lett.}\ }\textbf {\bibinfo {volume} {117}},\ \bibinfo {pages} {231601} (\bibinfo {year} {2016})}\BibitemShut {NoStop}%
\bibitem [{\citenamefont {Linde}(1981)}]{LINDE198137}%
  \BibitemOpen
  \bibfield  {author} {\bibinfo {author} {\bibfnamefont {A.}~\bibnamefont {Linde}},\ }\href {\doibase https://doi.org/10.1016/0370-2693(81)90281-1} {\bibfield  {journal} {\bibinfo  {journal} {Physics Letters B}\ }\textbf {\bibinfo {volume} {100}},\ \bibinfo {pages} {37} (\bibinfo {year} {1981})}\BibitemShut {NoStop}%
\bibitem [{\citenamefont {Linde}(1983)}]{Linde:1981zj}%
  \BibitemOpen
  \bibfield  {author} {\bibinfo {author} {\bibfnamefont {A.~D.}\ \bibnamefont {Linde}},\ }\href {\doibase 10.1016/0550-3213(83)90072-X} {\bibfield  {journal} {\bibinfo  {journal} {Nucl. Phys. B}\ }\textbf {\bibinfo {volume} {216}},\ \bibinfo {pages} {421} (\bibinfo {year} {1983})},\ \bibinfo {note} {[Erratum: Nucl.Phys.B 223, 544 (1983)]}\BibitemShut {NoStop}%
\bibitem [{\citenamefont {Coleman}\ and\ \citenamefont {De~Luccia}(1980)}]{PhysRevD.21.3305}%
  \BibitemOpen
  \bibfield  {author} {\bibinfo {author} {\bibfnamefont {S.}~\bibnamefont {Coleman}}\ and\ \bibinfo {author} {\bibfnamefont {F.}~\bibnamefont {De~Luccia}},\ }\href {\doibase 10.1103/PhysRevD.21.3305} {\bibfield  {journal} {\bibinfo  {journal} {Phys. Rev. D}\ }\textbf {\bibinfo {volume} {21}},\ \bibinfo {pages} {3305} (\bibinfo {year} {1980})}\BibitemShut {NoStop}%
\bibitem [{\citenamefont {Hawking}(1971)}]{Hawking:1971ei}%
  \BibitemOpen
  \bibfield  {author} {\bibinfo {author} {\bibfnamefont {S.}~\bibnamefont {Hawking}},\ }\href {\doibase 10.1093/mnras/152.1.75} {\bibfield  {journal} {\bibinfo  {journal} {Mon. Not. Roy. Astron. Soc.}\ }\textbf {\bibinfo {volume} {152}},\ \bibinfo {pages} {75} (\bibinfo {year} {1971})}\BibitemShut {NoStop}%
\bibitem [{\citenamefont {Carr}\ and\ \citenamefont {Hawking}(1974)}]{Carr:1974nx}%
  \BibitemOpen
  \bibfield  {author} {\bibinfo {author} {\bibfnamefont {B.~J.}\ \bibnamefont {Carr}}\ and\ \bibinfo {author} {\bibfnamefont {S.~W.}\ \bibnamefont {Hawking}},\ }\href {\doibase 10.1093/mnras/168.2.399} {\bibfield  {journal} {\bibinfo  {journal} {Mon. Not. Roy. Astron. Soc.}\ }\textbf {\bibinfo {volume} {168}},\ \bibinfo {pages} {399} (\bibinfo {year} {1974})}\BibitemShut {NoStop}%
\bibitem [{\citenamefont {Carr}(1975)}]{Carr:1975qj}%
  \BibitemOpen
  \bibfield  {author} {\bibinfo {author} {\bibfnamefont {B.~J.}\ \bibnamefont {Carr}},\ }\href {\doibase 10.1086/153853} {\bibfield  {journal} {\bibinfo  {journal} {Astrophys. J.}\ }\textbf {\bibinfo {volume} {201}},\ \bibinfo {pages} {1} (\bibinfo {year} {1975})}\BibitemShut {NoStop}%
\bibitem [{\citenamefont {Berezin}\ \emph {et~al.}(1988)\citenamefont {Berezin}, \citenamefont {Kuzmin},\ and\ \citenamefont {Tkachev}}]{BEREZIN1988397}%
  \BibitemOpen
  \bibfield  {author} {\bibinfo {author} {\bibfnamefont {V.}~\bibnamefont {Berezin}}, \bibinfo {author} {\bibfnamefont {V.}~\bibnamefont {Kuzmin}}, \ and\ \bibinfo {author} {\bibfnamefont {I.}~\bibnamefont {Tkachev}},\ }\href {\doibase https://doi.org/10.1016/0370-2693(88)90672-7} {\bibfield  {journal} {\bibinfo  {journal} {Physics Letters B}\ }\textbf {\bibinfo {volume} {207}},\ \bibinfo {pages} {397} (\bibinfo {year} {1988})}\BibitemShut {NoStop}%
\bibitem [{\citenamefont {Oshita}\ \emph {et~al.}(2019)\citenamefont {Oshita}, \citenamefont {Yamada},\ and\ \citenamefont {Yamaguchi}}]{Oshita:2018ptr}%
  \BibitemOpen
  \bibfield  {author} {\bibinfo {author} {\bibfnamefont {N.}~\bibnamefont {Oshita}}, \bibinfo {author} {\bibfnamefont {M.}~\bibnamefont {Yamada}}, \ and\ \bibinfo {author} {\bibfnamefont {M.}~\bibnamefont {Yamaguchi}},\ }\href {\doibase 10.1016/j.physletb.2019.02.032} {\bibfield  {journal} {\bibinfo  {journal} {Phys. Lett. B}\ }\textbf {\bibinfo {volume} {791}},\ \bibinfo {pages} {149} (\bibinfo {year} {2019})},\ \Eprint {http://arxiv.org/abs/1808.01382} {arXiv:1808.01382 [gr-qc]} \BibitemShut {NoStop}%
\bibitem [{\citenamefont {Hiscock}(1996)}]{Hiscock:1995ma}%
  \BibitemOpen
  \bibfield  {author} {\bibinfo {author} {\bibfnamefont {W.~A.}\ \bibnamefont {Hiscock}},\ }\href {\doibase 10.1016/0370-2693(95)01345-8} {\bibfield  {journal} {\bibinfo  {journal} {Phys. Lett. B}\ }\textbf {\bibinfo {volume} {366}},\ \bibinfo {pages} {77} (\bibinfo {year} {1996})},\ \Eprint {http://arxiv.org/abs/gr-qc/9510003} {arXiv:gr-qc/9510003} \BibitemShut {NoStop}%
\bibitem [{\citenamefont {Preskill}\ and\ \citenamefont {Vilenkin}(1993)}]{PhysRevD.47.2324}%
  \BibitemOpen
  \bibfield  {author} {\bibinfo {author} {\bibfnamefont {J.}~\bibnamefont {Preskill}}\ and\ \bibinfo {author} {\bibfnamefont {A.}~\bibnamefont {Vilenkin}},\ }\href {\doibase 10.1103/PhysRevD.47.2324} {\bibfield  {journal} {\bibinfo  {journal} {Phys. Rev. D}\ }\textbf {\bibinfo {volume} {47}},\ \bibinfo {pages} {2324} (\bibinfo {year} {1993})}\BibitemShut {NoStop}%
\bibitem [{\citenamefont {Dasgupta}(1997)}]{Dasgupta:1997kn}%
  \BibitemOpen
  \bibfield  {author} {\bibinfo {author} {\bibfnamefont {I.}~\bibnamefont {Dasgupta}},\ }\href {\doibase 10.1016/S0550-3213(97)00546-4} {\bibfield  {journal} {\bibinfo  {journal} {Nucl. Phys. B}\ }\textbf {\bibinfo {volume} {506}},\ \bibinfo {pages} {421} (\bibinfo {year} {1997})},\ \Eprint {http://arxiv.org/abs/hep-th/9702041} {arXiv:hep-th/9702041} \BibitemShut {NoStop}%
\bibitem [{\citenamefont {Yajnik}\ and\ \citenamefont {Padmanabhan}(1987)}]{PhysRevD.35.3100}%
  \BibitemOpen
  \bibfield  {author} {\bibinfo {author} {\bibfnamefont {U.~A.}\ \bibnamefont {Yajnik}}\ and\ \bibinfo {author} {\bibfnamefont {T.}~\bibnamefont {Padmanabhan}},\ }\href {\doibase 10.1103/PhysRevD.35.3100} {\bibfield  {journal} {\bibinfo  {journal} {Phys. Rev. D}\ }\textbf {\bibinfo {volume} {35}},\ \bibinfo {pages} {3100} (\bibinfo {year} {1987})}\BibitemShut {NoStop}%
\bibitem [{\citenamefont {Yajnik}(1986)}]{PhysRevD.34.1237}%
  \BibitemOpen
  \bibfield  {author} {\bibinfo {author} {\bibfnamefont {U.~A.}\ \bibnamefont {Yajnik}},\ }\href {\doibase 10.1103/PhysRevD.34.1237} {\bibfield  {journal} {\bibinfo  {journal} {Phys. Rev. D}\ }\textbf {\bibinfo {volume} {34}},\ \bibinfo {pages} {1237} (\bibinfo {year} {1986})}\BibitemShut {NoStop}%
\bibitem [{\citenamefont {Kumar}\ and\ \citenamefont {Yajnik}(2009)}]{Kumar:2008jb}%
  \BibitemOpen
  \bibfield  {author} {\bibinfo {author} {\bibfnamefont {B.}~\bibnamefont {Kumar}}\ and\ \bibinfo {author} {\bibfnamefont {U.~A.}\ \bibnamefont {Yajnik}},\ }\href {\doibase 10.1103/PhysRevD.79.065001} {\bibfield  {journal} {\bibinfo  {journal} {Phys. Rev. D}\ }\textbf {\bibinfo {volume} {79}},\ \bibinfo {pages} {065001} (\bibinfo {year} {2009})},\ \Eprint {http://arxiv.org/abs/0807.3254} {arXiv:0807.3254 [hep-th]} \BibitemShut {NoStop}%
\bibitem [{\citenamefont {Firouzjahi}\ \emph {et~al.}(2020)\citenamefont {Firouzjahi}, \citenamefont {Karami},\ and\ \citenamefont {Rostami}}]{Firouzjahi:2020hfq}%
  \BibitemOpen
  \bibfield  {author} {\bibinfo {author} {\bibfnamefont {H.}~\bibnamefont {Firouzjahi}}, \bibinfo {author} {\bibfnamefont {A.}~\bibnamefont {Karami}}, \ and\ \bibinfo {author} {\bibfnamefont {T.}~\bibnamefont {Rostami}},\ }\href {\doibase 10.1103/PhysRevD.101.104036} {\bibfield  {journal} {\bibinfo  {journal} {Phys. Rev. D}\ }\textbf {\bibinfo {volume} {101}},\ \bibinfo {pages} {104036} (\bibinfo {year} {2020})},\ \Eprint {http://arxiv.org/abs/2002.04856} {arXiv:2002.04856 [gr-qc]} \BibitemShut {NoStop}%
\bibitem [{\citenamefont {Steinhardt}(1981)}]{Steinhardt:1981ec}%
  \BibitemOpen
  \bibfield  {author} {\bibinfo {author} {\bibfnamefont {P.~J.}\ \bibnamefont {Steinhardt}},\ }\href {\doibase 10.1016/0550-3213(81)90449-1} {\bibfield  {journal} {\bibinfo  {journal} {Nucl. Phys. B}\ }\textbf {\bibinfo {volume} {190}},\ \bibinfo {pages} {583} (\bibinfo {year} {1981})}\BibitemShut {NoStop}%
\bibitem [{\citenamefont {Lee}\ \emph {et~al.}(2013)\citenamefont {Lee}, \citenamefont {Lee}, \citenamefont {MacKenzie}, \citenamefont {Paranjape}, \citenamefont {Yajnik},\ and\ \citenamefont {Yeom}}]{Lee:2013ega}%
  \BibitemOpen
  \bibfield  {author} {\bibinfo {author} {\bibfnamefont {B.-H.}\ \bibnamefont {Lee}}, \bibinfo {author} {\bibfnamefont {W.}~\bibnamefont {Lee}}, \bibinfo {author} {\bibfnamefont {R.}~\bibnamefont {MacKenzie}}, \bibinfo {author} {\bibfnamefont {M.~B.}\ \bibnamefont {Paranjape}}, \bibinfo {author} {\bibfnamefont {U.~A.}\ \bibnamefont {Yajnik}}, \ and\ \bibinfo {author} {\bibfnamefont {D.-h.}\ \bibnamefont {Yeom}},\ }\href {\doibase 10.1103/PhysRevD.88.085031} {\bibfield  {journal} {\bibinfo  {journal} {Phys. Rev. D}\ }\textbf {\bibinfo {volume} {88}},\ \bibinfo {pages} {085031} (\bibinfo {year} {2013})},\ \Eprint {http://arxiv.org/abs/1308.3501} {arXiv:1308.3501 [hep-th]} \BibitemShut {NoStop}%
\bibitem [{\citenamefont {Kumar}\ \emph {et~al.}(2010)\citenamefont {Kumar}, \citenamefont {Paranjape},\ and\ \citenamefont {Yajnik}}]{Kumar:2010mv}%
  \BibitemOpen
  \bibfield  {author} {\bibinfo {author} {\bibfnamefont {B.}~\bibnamefont {Kumar}}, \bibinfo {author} {\bibfnamefont {M.~B.}\ \bibnamefont {Paranjape}}, \ and\ \bibinfo {author} {\bibfnamefont {U.~A.}\ \bibnamefont {Yajnik}},\ }\href {\doibase 10.1103/PhysRevD.82.025022} {\bibfield  {journal} {\bibinfo  {journal} {Phys. Rev. D}\ }\textbf {\bibinfo {volume} {82}},\ \bibinfo {pages} {025022} (\bibinfo {year} {2010})},\ \Eprint {http://arxiv.org/abs/1006.0693} {arXiv:1006.0693 [hep-th]} \BibitemShut {NoStop}%
\bibitem [{\citenamefont {Blasi}\ and\ \citenamefont {Mariotti}(2022)}]{Blasi:2022woz}%
  \BibitemOpen
  \bibfield  {author} {\bibinfo {author} {\bibfnamefont {S.}~\bibnamefont {Blasi}}\ and\ \bibinfo {author} {\bibfnamefont {A.}~\bibnamefont {Mariotti}},\ }\href {\doibase 10.1103/PhysRevLett.129.261303} {\bibfield  {journal} {\bibinfo  {journal} {Phys. Rev. Lett.}\ }\textbf {\bibinfo {volume} {129}},\ \bibinfo {pages} {261303} (\bibinfo {year} {2022})},\ \Eprint {http://arxiv.org/abs/2203.16450} {arXiv:2203.16450 [hep-ph]} \BibitemShut {NoStop}%
\bibitem [{\citenamefont {Moss}(1985)}]{PhysRevD.32.1333}%
  \BibitemOpen
  \bibfield  {author} {\bibinfo {author} {\bibfnamefont {I.~G.}\ \bibnamefont {Moss}},\ }\href {\doibase 10.1103/PhysRevD.32.1333} {\bibfield  {journal} {\bibinfo  {journal} {Phys. Rev. D}\ }\textbf {\bibinfo {volume} {32}},\ \bibinfo {pages} {1333} (\bibinfo {year} {1985})}\BibitemShut {NoStop}%
\bibitem [{\citenamefont {Cheung}\ and\ \citenamefont {Leichenauer}(2014)}]{Cheung:2013sxa}%
  \BibitemOpen
  \bibfield  {author} {\bibinfo {author} {\bibfnamefont {C.}~\bibnamefont {Cheung}}\ and\ \bibinfo {author} {\bibfnamefont {S.}~\bibnamefont {Leichenauer}},\ }\href {\doibase 10.1103/PhysRevD.89.104035} {\bibfield  {journal} {\bibinfo  {journal} {Phys. Rev. D}\ }\textbf {\bibinfo {volume} {89}},\ \bibinfo {pages} {104035} (\bibinfo {year} {2014})},\ \Eprint {http://arxiv.org/abs/1309.0530} {arXiv:1309.0530 [hep-ph]} \BibitemShut {NoStop}%
\bibitem [{\citenamefont {Green}\ \emph {et~al.}(2006)\citenamefont {Green}, \citenamefont {Silverstein},\ and\ \citenamefont {Starr}}]{Green:2006nv}%
  \BibitemOpen
  \bibfield  {author} {\bibinfo {author} {\bibfnamefont {D.~R.}\ \bibnamefont {Green}}, \bibinfo {author} {\bibfnamefont {E.}~\bibnamefont {Silverstein}}, \ and\ \bibinfo {author} {\bibfnamefont {D.}~\bibnamefont {Starr}},\ }\href {\doibase 10.1103/PhysRevD.74.024004} {\bibfield  {journal} {\bibinfo  {journal} {Phys. Rev. D}\ }\textbf {\bibinfo {volume} {74}},\ \bibinfo {pages} {024004} (\bibinfo {year} {2006})},\ \Eprint {http://arxiv.org/abs/hep-th/0605047} {arXiv:hep-th/0605047} \BibitemShut {NoStop}%
\bibitem [{\citenamefont {Burda}\ \emph {et~al.}(2015)\citenamefont {Burda}, \citenamefont {Gregory},\ and\ \citenamefont {Moss}}]{Burda:2015yfa}%
  \BibitemOpen
  \bibfield  {author} {\bibinfo {author} {\bibfnamefont {P.}~\bibnamefont {Burda}}, \bibinfo {author} {\bibfnamefont {R.}~\bibnamefont {Gregory}}, \ and\ \bibinfo {author} {\bibfnamefont {I.}~\bibnamefont {Moss}},\ }\href {\doibase 10.1007/JHEP08(2015)114} {\bibfield  {journal} {\bibinfo  {journal} {JHEP}\ }\textbf {\bibinfo {volume} {08}},\ \bibinfo {pages} {114} (\bibinfo {year} {2015})},\ \Eprint {http://arxiv.org/abs/1503.07331} {arXiv:1503.07331 [hep-th]} \BibitemShut {NoStop}%
\bibitem [{\citenamefont {Mukaida}\ and\ \citenamefont {Yamada}(2017)}]{Mukaida:2017bgd}%
  \BibitemOpen
  \bibfield  {author} {\bibinfo {author} {\bibfnamefont {K.}~\bibnamefont {Mukaida}}\ and\ \bibinfo {author} {\bibfnamefont {M.}~\bibnamefont {Yamada}},\ }\href {\doibase 10.1103/PhysRevD.96.103514} {\bibfield  {journal} {\bibinfo  {journal} {Phys. Rev. D}\ }\textbf {\bibinfo {volume} {96}},\ \bibinfo {pages} {103514} (\bibinfo {year} {2017})},\ \Eprint {http://arxiv.org/abs/1706.04523} {arXiv:1706.04523 [hep-th]} \BibitemShut {NoStop}%
\bibitem [{\citenamefont {El-Menoufi}\ \emph {et~al.}(2020)\citenamefont {El-Menoufi}, \citenamefont {Huber},\ and\ \citenamefont {Manuel}}]{El-Menoufi:2020ron}%
  \BibitemOpen
  \bibfield  {author} {\bibinfo {author} {\bibfnamefont {B.~K.}\ \bibnamefont {El-Menoufi}}, \bibinfo {author} {\bibfnamefont {S.~J.}\ \bibnamefont {Huber}}, \ and\ \bibinfo {author} {\bibfnamefont {J.~P.}\ \bibnamefont {Manuel}},\ }\href@noop {} {\  (\bibinfo {year} {2020})},\ \Eprint {http://arxiv.org/abs/2006.16275} {arXiv:2006.16275 [hep-th]} \BibitemShut {NoStop}%
\bibitem [{\citenamefont {Hiscock}(1987)}]{PhysRevD.35.1161}%
  \BibitemOpen
  \bibfield  {author} {\bibinfo {author} {\bibfnamefont {W.~A.}\ \bibnamefont {Hiscock}},\ }\href {\doibase 10.1103/PhysRevD.35.1161} {\bibfield  {journal} {\bibinfo  {journal} {Phys. Rev. D}\ }\textbf {\bibinfo {volume} {35}},\ \bibinfo {pages} {1161} (\bibinfo {year} {1987})}\BibitemShut {NoStop}%
\bibitem [{\citenamefont {Ai}(2019)}]{Ai:2018rnh}%
  \BibitemOpen
  \bibfield  {author} {\bibinfo {author} {\bibfnamefont {W.-Y.}\ \bibnamefont {Ai}},\ }\href {\doibase 10.1007/JHEP03(2019)164} {\bibfield  {journal} {\bibinfo  {journal} {JHEP}\ }\textbf {\bibinfo {volume} {03}},\ \bibinfo {pages} {164} (\bibinfo {year} {2019})},\ \Eprint {http://arxiv.org/abs/1812.06962} {arXiv:1812.06962 [hep-th]} \BibitemShut {NoStop}%
\bibitem [{\citenamefont {Briaud}\ \emph {et~al.}(2022)\citenamefont {Briaud}, \citenamefont {Shkerin},\ and\ \citenamefont {Sibiryakov}}]{PhysRevD.106.125001}%
  \BibitemOpen
  \bibfield  {author} {\bibinfo {author} {\bibfnamefont {V.}~\bibnamefont {Briaud}}, \bibinfo {author} {\bibfnamefont {A.}~\bibnamefont {Shkerin}}, \ and\ \bibinfo {author} {\bibfnamefont {S.}~\bibnamefont {Sibiryakov}},\ }\href {\doibase 10.1103/PhysRevD.106.125001} {\bibfield  {journal} {\bibinfo  {journal} {Phys. Rev. D}\ }\textbf {\bibinfo {volume} {106}},\ \bibinfo {pages} {125001} (\bibinfo {year} {2022})}\BibitemShut {NoStop}%
\bibitem [{\citenamefont {Gregory}\ \emph {et~al.}(2014)\citenamefont {Gregory}, \citenamefont {Moss},\ and\ \citenamefont {Withers}}]{Gregory:2013hja}%
  \BibitemOpen
  \bibfield  {author} {\bibinfo {author} {\bibfnamefont {R.}~\bibnamefont {Gregory}}, \bibinfo {author} {\bibfnamefont {I.~G.}\ \bibnamefont {Moss}}, \ and\ \bibinfo {author} {\bibfnamefont {B.}~\bibnamefont {Withers}},\ }\href {\doibase 10.1007/JHEP03(2014)081} {\bibfield  {journal} {\bibinfo  {journal} {JHEP}\ }\textbf {\bibinfo {volume} {03}},\ \bibinfo {pages} {081} (\bibinfo {year} {2014})},\ \Eprint {http://arxiv.org/abs/1401.0017} {arXiv:1401.0017 [hep-th]} \BibitemShut {NoStop}%
\bibitem [{\citenamefont {Miyachi}\ and\ \citenamefont {Soda}(2021)}]{Miyachi:2021bwd}%
  \BibitemOpen
  \bibfield  {author} {\bibinfo {author} {\bibfnamefont {T.}~\bibnamefont {Miyachi}}\ and\ \bibinfo {author} {\bibfnamefont {J.}~\bibnamefont {Soda}},\ }\href {\doibase 10.1103/PhysRevD.103.085009} {\bibfield  {journal} {\bibinfo  {journal} {Phys. Rev. D}\ }\textbf {\bibinfo {volume} {103}},\ \bibinfo {pages} {085009} (\bibinfo {year} {2021})},\ \Eprint {http://arxiv.org/abs/2102.02462} {arXiv:2102.02462 [gr-qc]} \BibitemShut {NoStop}%
\bibitem [{\citenamefont {Gibbons}\ and\ \citenamefont {Hawking}(1977)}]{Gibbons:1976ue}%
  \BibitemOpen
  \bibfield  {author} {\bibinfo {author} {\bibfnamefont {G.~W.}\ \bibnamefont {Gibbons}}\ and\ \bibinfo {author} {\bibfnamefont {S.~W.}\ \bibnamefont {Hawking}},\ }\href {\doibase 10.1103/PhysRevD.15.2752} {\bibfield  {journal} {\bibinfo  {journal} {Phys. Rev. D}\ }\textbf {\bibinfo {volume} {15}},\ \bibinfo {pages} {2752} (\bibinfo {year} {1977})}\BibitemShut {NoStop}%
\bibitem [{\citenamefont {Wang}\ \emph {et~al.}(2020)\citenamefont {Wang}, \citenamefont {Huang},\ and\ \citenamefont {Zhang}}]{Wang:2020jsg}%
  \BibitemOpen
  \bibfield  {author} {\bibinfo {author} {\bibfnamefont {X.}~\bibnamefont {Wang}}, \bibinfo {author} {\bibfnamefont {F.~P.}\ \bibnamefont {Huang}}, \ and\ \bibinfo {author} {\bibfnamefont {X.}~\bibnamefont {Zhang}},\ }\href {\doibase 10.1088/1475-7516/2020/05/045} {\bibfield  {journal} {\bibinfo  {journal} {JCAP}\ }\textbf {\bibinfo {volume} {05}},\ \bibinfo {pages} {045} (\bibinfo {year} {2020})},\ \Eprint {http://arxiv.org/abs/2003.08892} {arXiv:2003.08892 [hep-ph]} \BibitemShut {NoStop}%
\bibitem [{\citenamefont {Goldstone}\ and\ \citenamefont {Jackiw}(1975)}]{PhysRevD.11.1486}%
  \BibitemOpen
  \bibfield  {author} {\bibinfo {author} {\bibfnamefont {J.}~\bibnamefont {Goldstone}}\ and\ \bibinfo {author} {\bibfnamefont {R.}~\bibnamefont {Jackiw}},\ }\href {\doibase 10.1103/PhysRevD.11.1486} {\bibfield  {journal} {\bibinfo  {journal} {Phys. Rev. D}\ }\textbf {\bibinfo {volume} {11}},\ \bibinfo {pages} {1486} (\bibinfo {year} {1975})}\BibitemShut {NoStop}%
\end{thebibliography}%

\end{document}